\documentclass[twocolumn]{aastex63}

\graphicspath{{./}{figures/}}
\usepackage{pifont}
\usepackage{multirow}
\usepackage{amsmath}

\begin{document}

\title{Kilonova and Optical Afterglow from Binary Neutron Star Mergers. I. Luminosity Function and Color Evolution}

\author[0000-0002-9195-4904]{Jin-Ping Zhu}
\affil{Department of Astronomy, School of Physics, Peking University, Beijing 100871, China; \url{zhujp@pku.edu.cn}}

\author[0000-0001-6374-8313]{Yuan-Pei Yang}
\affiliation{South-Western Institute for Astronomy Research, Yunnan University, Kunming, Yunnan 650500, People’s Republic of China; \url{ypyang@ynu.edu.cn}}

\collaboration{6}{(these authors contributed equally to this work)}

\author[0000-0002-9725-2524]{Bing Zhang}
\affiliation{Nevada Center for Astrophysics, University of Nevada, Las Vegas, NV 89154, USA; \url{bing.zhang@unlv.edu}}
\affiliation{Department of Physics and Astronomy, University of Nevada, Las Vegas, NV 89154, USA}

\author[0000-0002-3100-6558]{He Gao}
\affiliation{Department of Astronomy, Beijing Normal University, Beijing 100875, China}

\author[0000-0002-1067-1911]{Yun-Wei Yu}
\affiliation{Institute of Astrophysics, Central China Normal University, Wuhan 430079, China}

\begin{abstract}

In the first work of this series, we adopt a GW170817-like, viewing-angle-dependent kilonova model and the standard afterglow model with the lightcurve distribution based on the properties of cosmological short gamma-ray bursts afterglows to simulate the luminosity functions and color evolution of both kilonovae and optical afterglow emissions from binary neutron star (BNS) mergers. {We find that $\sim10\%$ afterglows are brighter than the associated kilonovae at the peak time, most of which are on-axis or nearly-on-axis. These kilonovae would be significantly polluted by the associated afterglow emission. Only at large viewing angles with $\sin\theta_{\rm v}\gtrsim 0.20$, the EM signals of most BNS mergers would be kilonova-dominated and some off-axis afterglows may emerge at $\sim5-10$\,day after the mergers.} At brightness dimmer than $\sim23-24$\,mag, according to their luminosity functions, the number of afterglows is much larger than that of kilonovae. Since the search depth of the present survey projects is $<22$\,mag, the number of afterglow events detected via serendipitous observations would be much higher than that of kilonova events, consistent with the current observations. {For the foreseeable survey projects (e.g., Mephisto, WFST, LSST) whose search depths can reach $\gtrsim23-24$\,mag, the detection rate of kilonovae could have the same order of magnitude as that of afterglows.} We also find that it may be difficult to use the fading rate in a single band to directly identify kilonovae and afterglows among various fast-evolving transients by serendipitous surveys. However, the color evolution between optical and infrared bands can identify them, since their color evolution patterns are unique compared with those of other fast-evolving transients.

\end{abstract}


\section{Introduction} 

Binary neutron star (BNS) and neutron star--black hole (NSBH) mergers are the prime gravitational waves (GW) sources for the current and future ground-based GW detectors \citep{abbott2020prospects}. They have been proposed to be the progenitors of short gamma-ray bursts \citep[sGRBs;][]{paczynski1986,paczynski1991,eichler1989,narayan1992,zhangb2018}, their associated broad-band afterglows\footnote{Besides short GRB afterglows, if a BNS merger leaves behind a long-lived highly-magnetized NS, the dissipation of the magnetar wind can power an early X-ray afterglow \citep{zhang2013,sun2017,sun2019,xiao2019,xue2019,ai2021}. The relativistic magnetar-powered ejecta can interact with the ambient medium to drive a bright late-time afterglow \citep{gao2013,liu2020}.}, and kilonova emission. The merger of BNS or NSBH would drive a ultra-relativistic jet, which may be powered by accretion of a massive remnant disk onto the newly formed BH or NS after the merger \citep{rezzolla2011,paschalidis2015,ruiz2016}. The interaction of the relativistic jet with the ambient medium generates multiwavelength nonthermal afterglow emission ranging from radio to X-rays \citep{rees1992,meszaros1993,paczynski1993,meszaros1997,sari1998}. Since the radiation of afterglow is beamed, if the viewing angle is far outside the core of the relativistic jet, the afterglow would have a low peak luminosity and with a delayed lightcurve rise. In addition to the beamed sGRB and its afterglow, an amount of neutron-rich matter is approximately isotropically released, which undergoes rapid neutron capture ($r$-process) nucleosynthesis  \citep{lattimer1974,lattimer1976,symbalisty1982}. The radioactive decay of $r$-process nuclei powers a rapidly evolving thermal kilonova in the ultraviolet-optical-infrared bands\footnote{Some BNS and NSBH mergers are predicted to occur in the active galactic nucleus (AGN) accretion disks \citep[e.g.,][]{cheng1999,mckernan2020}. sGRBs embedded in the AGN disks may be choked by the disk atmosphere \citep{zhu2021highenergy1,zhu2021neutron,perna2021}, with kilonova emission being outshone by disk emission \citep{zhu2021neutron}. AGN BNS and NSBH mergers could be an important neutrino source to contribute the astrophysical neutrino background \citep{zhu2021highenergy2,zhu2021highenergy1}.} \citep{li1998,metzger2010}.

The first BNS merger gravitational-wave (GW) event, GW170817 \citep{abbott2017GW170817}, was detected by the Laser Interferometer Gravitational-Wave Observatory (LIGO) \citep{aasi2015} and Virgo \citep{acernese2015}. At $\sim1.7\,{\rm s}$ after the merger, GRB170817A with a duration of $\sim2\,{\rm s}$ was triggered by the Fermi GRB Monitor \citep{abbott2017gravitational,goldstein2017,savchenko2017,zhangbb2018}, and an associated kilonova AT2017gfo was discovered in the host galaxy NGC\,4993 $\sim11$\,hr later \citep{abbott2017multimessenger,arcavi2017,coulter2017,drout2017,evans2017,kasliwal2017,pian2017,smartt2017,kilpatrick2017,lamb2019optical}. An off-axis multiwavelength afterglow (from radio to X-ray) was detected with luminosity peaking at $\sim200\,{\rm d}$ \citep{margutti2017,troja2017,lazzati2018,lyman2018,ghirlanda2019}. The multi-messenger observations of this event provided smoking-gun evidence for the BNS merger origin of sGRBs. \cite{abbott2021observation} recently reported the observations of GWs from two NSBH mergers. Despite many efforts for the follow-up observations, no confirmed kilonova or afterglow candidate were identified \citep[e.g.,][]{anand2021,coughlin2020b} which may suggest that the two NSBH merger events were plunging events without tidal disruption of the NSs \citep{abbott2021observation,zhu2021population,zhu2021no,gompertz2021}. In addition to AT2017gfo, some kilonova candidates have been claimed to be in the superposition with the decaying sGRB afterglows in the past few years \citep[e.g.,][]{berger2013,tanvir2013,fan2013,gao2015,gao2017,jin2015,jin2016,jin2020,yang2015,lamb2019,troja2019,ma2021,fong2021,wu2021,yuan2021}. Two main criteria were used to identify the kilonova candidates: 1.\,a significant excess in optical flux with respect to the sGRB optical afterglow lightcurve; 2.\,the existence of color evolution of  kilonova emission as compared with the non-evolving color of sGRB afterglow emission. However, for some candidates, the lack of  spectral or multi-color observations makes it difficult to discover color evolution, hindering the identifications of kilonova emission.

Observationally, the relatively low luminosity and fast-evolving nature of kilonovae not associated with sGRB afterglows makes it difficult to detect them. Without $\gamma$-ray triggers to launch target-of-opportunity observations, these transients have to be detected serendipitously. The properties of the electromagnetic (EM) counterparts of GW, kilonovae and afterglows can be predicted from theory, and the observed event rates would mainly depend on the observational depth and the sky area. Also the number of re-visits per field in a given time frame as well as the survey filters all play a role in efficiently identifying them \citep[e.g.,][]{scolnic2018,saguescarracedo2021,almualla2021,frostig2021,zhu2021kilonova,andreoni2021b}. Some kilonova candidates might have been recorded in the past optical survey projects. Thanks to the recently improved technology of wide-field surveys {\citep[e.g., Pan-STARRS and the Zwicky Transient Facility abbreviated as ZTF;][]{morgan2012,graham2019}}, identifying rapidly fading transients become more and more efficient. {Some fast-evolving transients, which had light curves lasting days to weeks and brightness in the range of theoretically predicted kilonova brightness, have been discovered \citep[e.g.,][]{drout2014,prentice2018,rest2018,mcbrien2021}. However, none of these events have been firmly identified as kilonovae.  \cite{andreoni2021} reported a few independent, optically-discovered GRB afterglows without any detection of kilonova candidates.}

In this series including two papers, we study the observation properties and survey strategies of optical counterparts of BNS mergers. In the first paper of the series, we study the observation properties of kilonova and afterglow of BNS mergers, including luminosity functions, viewing-angle-dependent lightcurves, and color evolution. In the second paper of the series\defcitealias{zhu2021kilonovaafterglow}{Paper II} \citep[][\citetalias{zhu2021kilonovaafterglow}]{zhu2021kilonovaafterglow}, based on the results of the first paper, we study how likely the kilonovae and afterglows from BNS mergers can be discovered in the future observations, and what would be the best strategy to discover them for wide-field surveys. This paper is organized as follows. We discuss the properties of kilonovae and afterglows, and constrain the physical parameters by the observations so far in Section \ref{sec:2}. We simulate luminosity distributions of kilonovae and afterglows, and discuss the brightness ratio between them at different bands in Section \ref{sec:3}. In Section \ref{sec:4}, we focus on color evolution properties of kilonovae, and compare them with other well-known fast-evolving transients. The color evolution of kilonova-contaminated afterglows is also analyzed. We summarize our results with discussion in Section \ref{sec:5}.

\section{Modeling} \label{sec:2}

In order to simulate the luminosity functions of the kilonovae and afterglows of BNS mergers, and the detectablility of optical EM counterparts from BNS mergers (\citetalias{zhu2021kilonovaafterglow}), one needs to know the redshift distribution of BNS mergers in the universe and cosmological radiation properties for both kilonovae and optical afterglows. 

\subsection{Redshift Distribution}

In principle, the observed system parameter and redshift distributions for BNS mergers should be the results of the convolution of both the intrinsic system parameters and redshift distribution. In this paper, we ignore possible redshift evolution of BNS system parameters \citep[e.g.,][]{sun2015} to allow us to separately discuss parameter distributions and redshift distribution $f(z)$ of BNS systems, where $z$ is the redshift. The number density per unit time for BNS mergers at a given redshift $z$ can be estimated as

\begin{equation}
\label{equ:merger_rate}
    \frac{d\dot{N}_{\rm BNS}}{dz} = \frac{\dot{\rho}_{0,{\rm BNS}}f(z)}{1 + z}\frac{dV(z)}{dz},
\end{equation}
where $\dot{\rho}_{0,{\rm BNS}}$ is the local BNS event rate density, and $f(z)$ is a dimensionless redshift distribution factor. The comoving volume element $dV(z)/dz$ in Equation (\ref{equ:merger_rate}) is

\begin{equation}
    \frac{dV}{dz} = \frac{c}{H_0}\frac{4\pi D_{\rm L}^2}{(1 + z)^2 \sqrt{\Omega_\Lambda + \Omega_{\rm m}(1 + z)^3}},
\end{equation}
where $c$ is the speed of light and $D_{\rm L}$ is the luminosity distance, which is expressed as

\begin{equation}
    D_{\rm L} = (1 + z) \frac{c}{H_0} \int_0^z \frac{dz}{\sqrt{\Omega_\Lambda + \Omega_{\rm m}(1 + z)^3}}
\end{equation}

BNS mergers would occur after a delay timescale with respect to the star formation history. Main types of delay-time distributions include the Gaussian delay model \citep{virgili2011}, log-normal delay model \citep{wanderman2015}, and power-law delay model \citep{virgili2011}. \cite{sun2015} suggested the log-normal delay model is one of the favored delay time models to explain the observations of sGRBs. Hereafter, we adopt the log-normal delay model as our merger delay model whose analytical fitting expression of $f(z)$ is adopted as Equation (A8) of \cite{zhu2021kilonova}.

\subsection{EM Emissions}
\subsubsection{Optical Afterglow Emission}

\begin{figure*}[tpb] 
    \centering
	\includegraphics[width = 0.48\linewidth , trim = 72 30 95 60, clip]{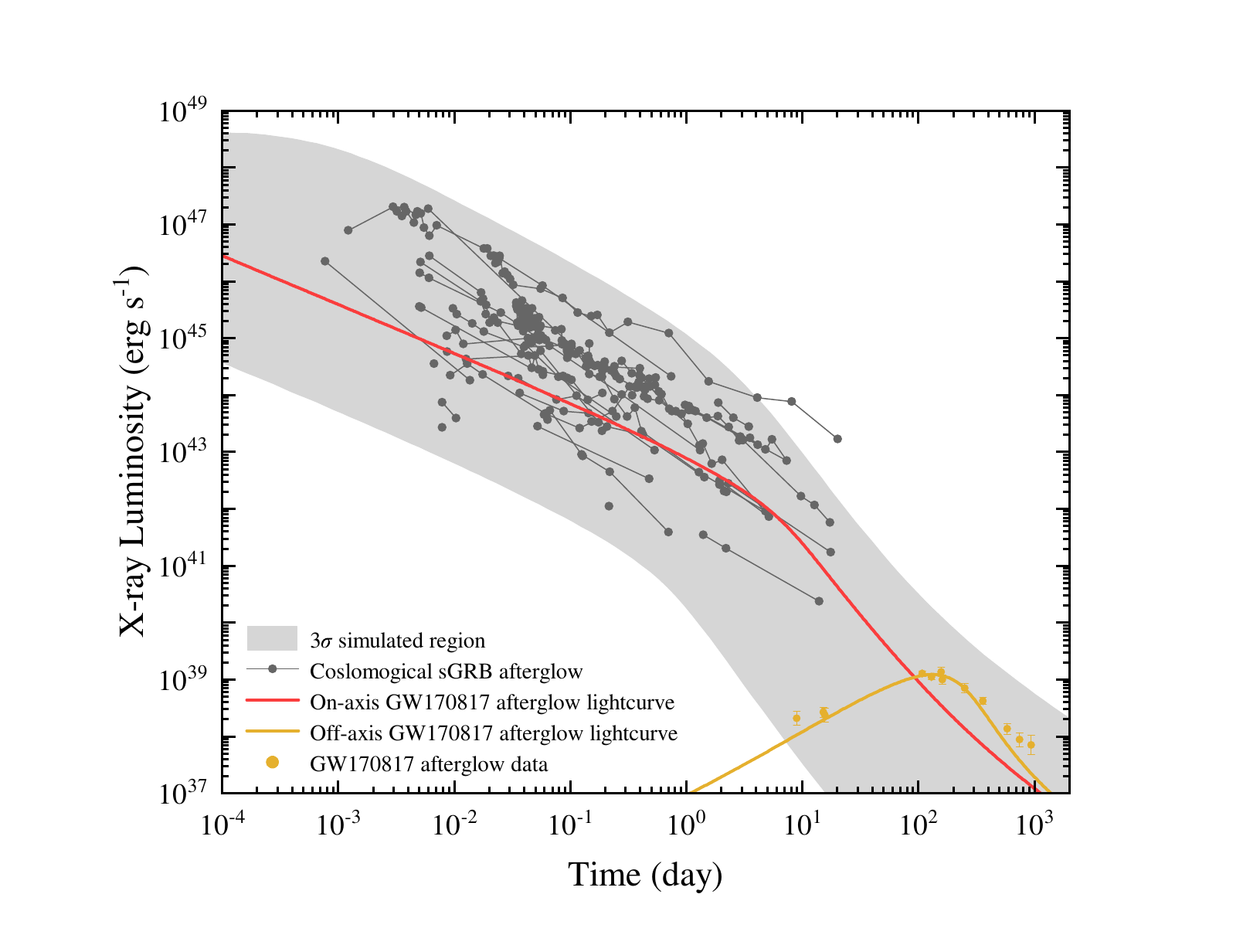}
	\includegraphics[width = 0.51\linewidth , trim = 72 30 60 60, clip]{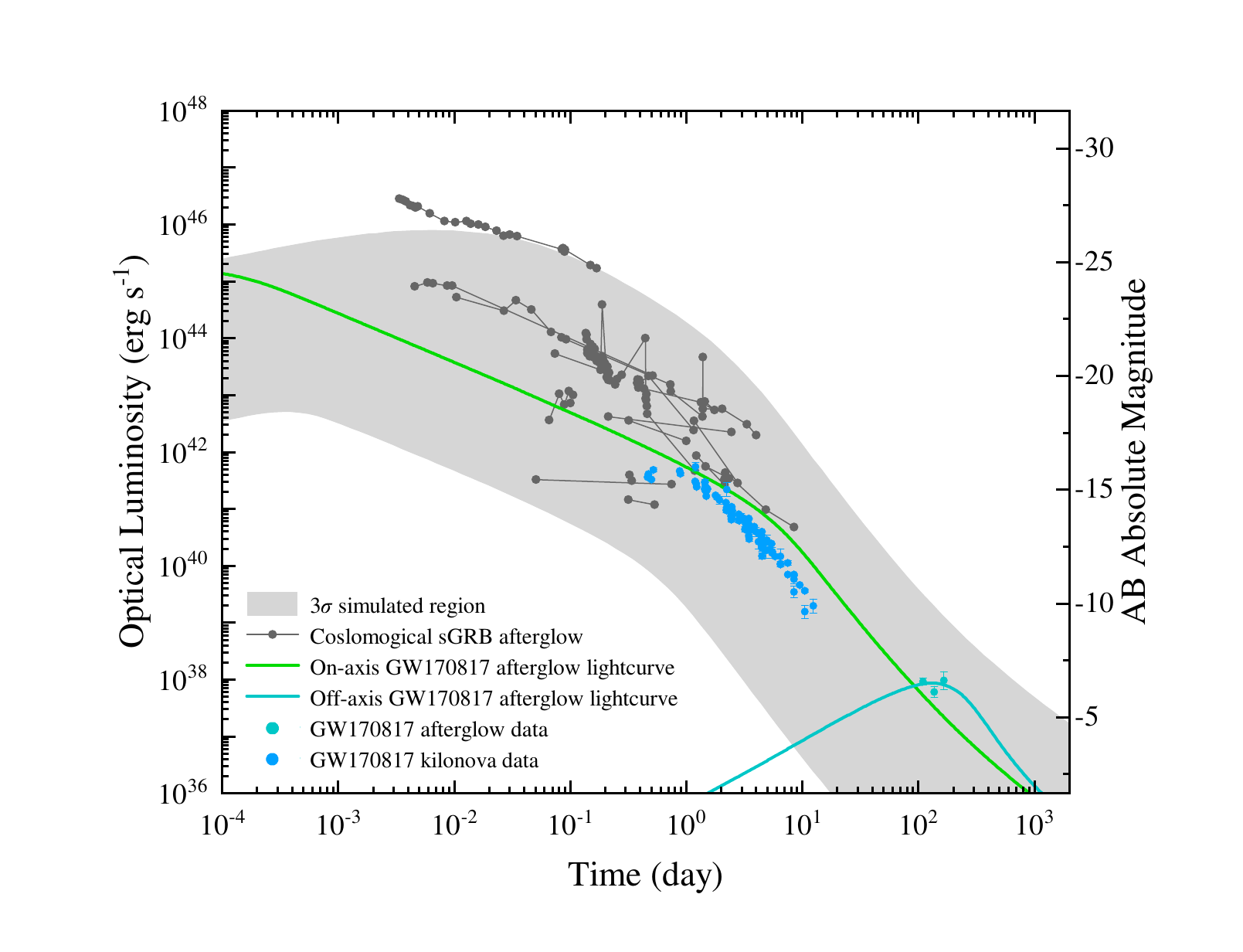}
	\includegraphics[width = 0.48\linewidth , trim = 72 30 95 60, clip]{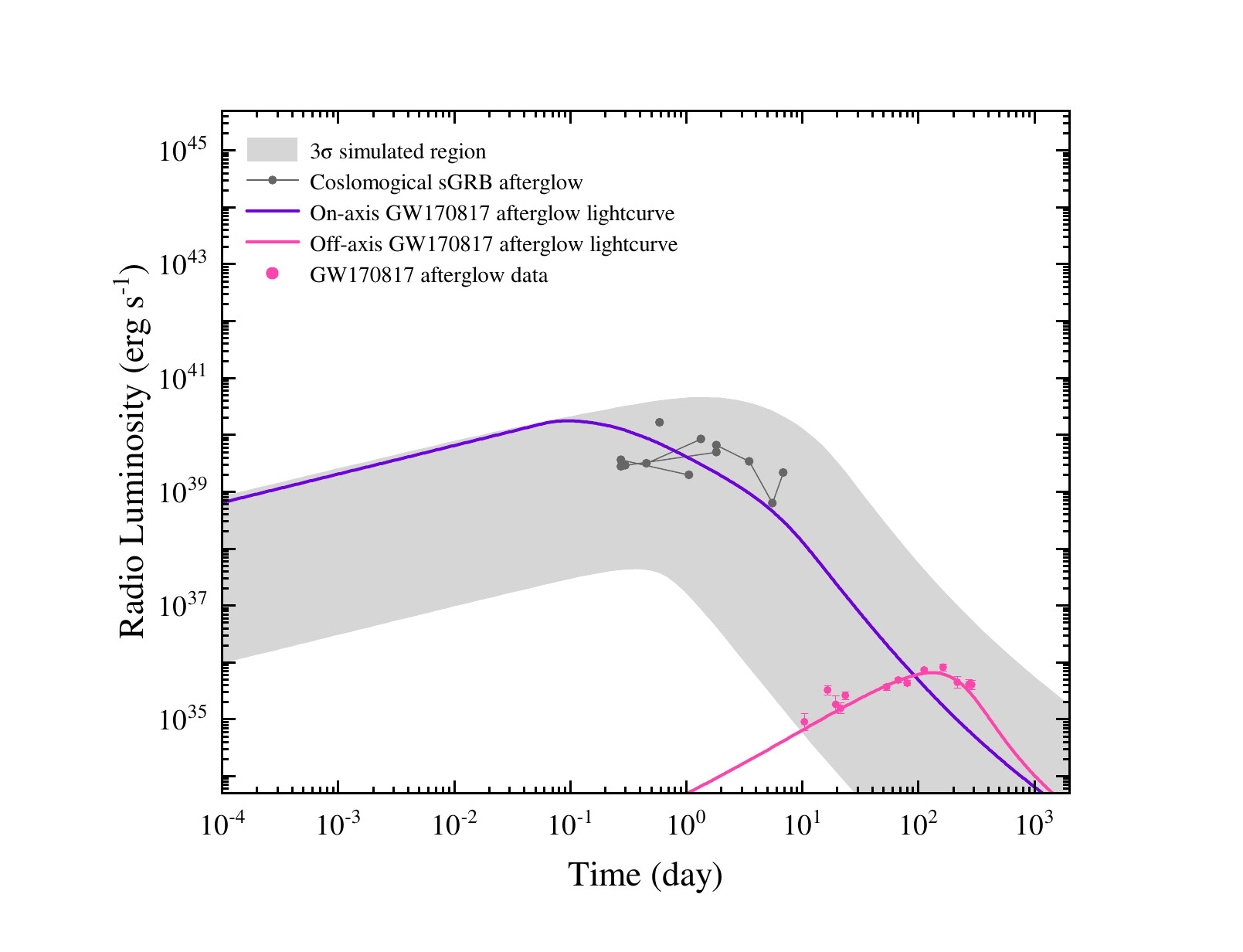}
    \caption{Rest-frame lightcurves of X-ray (top left panel), optical (top right  panel) and radio (bottom panel) afterglows. The gray points denote the afterglow data points of cosmological sGRB afterglows collected from \cite{fong2015} and \cite{rastinejad2021}. The gray region corresponds to the $3\sigma$ simulated region fitting the gray data points. The orange, cyan and magenta points and lines denote the observed afterglow data of GRB170817A and corresponding lightcurve fitting, respectively. The predicted on-axis X-ray, optical and radio afterglow lightcurves for GRB170817A are marked as red, green and purple lines, respectively. The blue points in the middle panel show the observed kilonova data of AT2017gfo \citep{villar2017}. Afterglow data were obtained from \cite{troja2017,troja2018,troja2020,piro2019,hallinan2017,lamb2019optical,lyman2018,resmi2018,margutti2018,mooley2018b}.}\label{fig1}
\end{figure*}

We first collect the sample of cosmological sGRB afterglows from \cite{fong2015} and \cite{rastinejad2021}, and constrain the afterglow physical parameters based on these afterglow samples. As shown in Figure \ref{fig1}, the gray points denote the rest-frame X-ray, optical, and radio lightcurves of the presently detected sGRB afterglows (not including GRB 170817A). In order to obtain simulated afterglow lightcurves, we briefly adopt the Gaussian structured jet model \citep[e.g.,][]{zhang2002} which was favored by the observations of GRB170817A afterglow \citep[e.g.,][]{alexander2017,lamb2018,lazzati2018,mooley2018b,troja2018,troja2020,ryan2020,xie2018,ghirlanda2019}, i.e.,

\begin{equation}
    E(\theta) = E_0\exp\left(-\frac{\theta^2}{2\theta_{\rm c}^2}\right),
\end{equation} 
where $E_0 = E_{\rm j} / (1 - \cos\theta_{\rm c})$ is the on-axis equivalent isotropic energy, $E_{\rm j}$ is the one-side jet energy, and $\theta_{\rm c}$ is the characteristic core angle. {We adopt a narrow jet structure, where we set $\theta_{\rm c} \sim 3^\circ$ based on recent simulations \citep[e.g.,][]{lamb2022} and observations inferred from the sGRB population \citep[e.g.,][]{jin2018,lamb2019}. } Furthermore, the spectra of the standard synchrotron emission from relativistic electrons are employed following \cite{sari1998,kumar2015,zhangb2018}. For more details of afterglow modeling we applied to calculate the sGRB lightcurves along the line of sight, see Appendix C in \cite{zhu2020}. We constrain the ranges of afterglow parameters, including $E_{\rm j}$, circumburst number density $n$, power-law index of the electron distribution $p$, fractions of shock energy carried by electrons $\varepsilon_e$ and by magnetic fields $\varepsilon_B$, to fit the cosmological multiwavelength afterglow lightcurves. We assume that each parameter is independent while the viewing angle $\theta_{\rm v}$ follows a random distribution between $0$ to $\theta_{\rm c}$. We present the $3\sigma$ simulated region shown as the gray regions in Figure \ref{fig1}. The simulated afterglow lightcurves can basically reproduce the observed cosmological sGRB afterglows in X-ray, optical, and radio bands. The distributions of these afterglow parameters, which are consistent with the constraints by \cite{fong2015,wang2015,wu2019,oconnor2020}, are listed in Table \ref{tab:1}. 

\begin{deluxetable}{cc}
\tablecaption{Afterglow Parameters Distribution Using Gaussian Jet Model \label{tab:1}}
\tablecolumns{2}
\tablewidth{0pt}
\tablehead{
\colhead{Parameter} &
\colhead{Distribution} 
}
\startdata
$E_{\rm j}$ & $\log_{10}E_{\rm j}/{\rm erg}\sim\mathcal{N}(49.3,0.4^2)$ \\
$\theta_{\rm c}$ & $\theta_{\rm c}/{\rm deg}\sim3$ \\
$p$ & $p\sim\mathcal{N}(2.25,0.1^2)$ \\
$n$ & $\log_{10}n/{\rm cm}^{-3}\sim\mathcal{N}(-2,0.4^2)$ \\
$\varepsilon_e$ & $\log_{10}\varepsilon_e\sim\mathcal{N}(-1,0.3^2)$ \\
$\varepsilon_B$ & $\log_{10}\varepsilon_B\sim\mathcal{N}(-3,0.4^2)$
\enddata
\tablecomments{$\mathcal{N}$ represents normal distribution.}
\end{deluxetable}

\subsubsection{Kilonova Emission}

\begin{figure}[tpb] 
    \centering
	\includegraphics[width = 1\linewidth , trim = 72 30 50 60, clip]{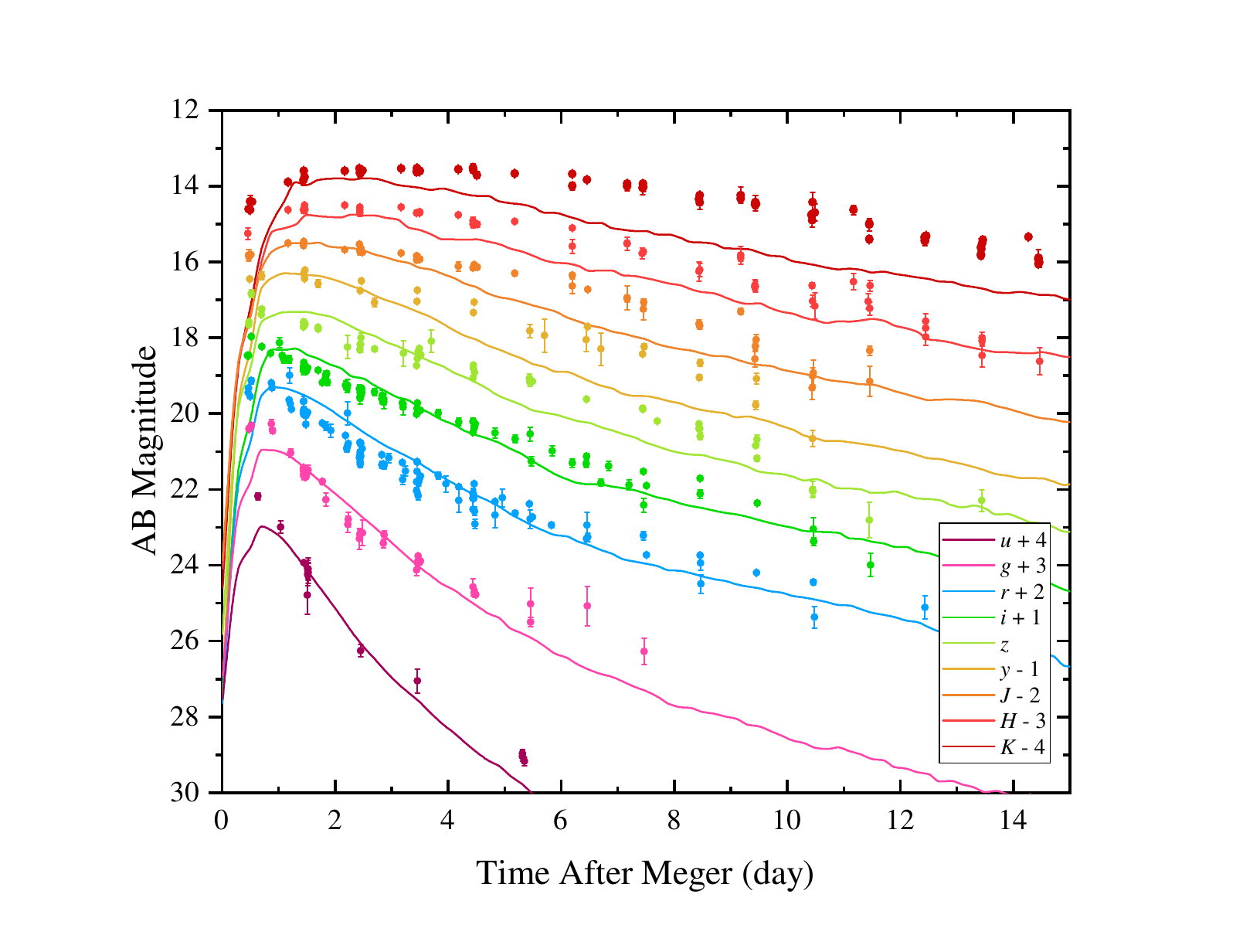}
    \caption{Multiwavelength observed kilonova data (colored points) of AT2017gfo and our fitting model (colored lines). Kilonova data were collected from \cite{andreoni2017,arcavi2017,coulter2017,cowperthwaite2017,diaz2017,drout2017,evans2017,hu2017,valenti2017,pozanenko2018,shappee2017,smartt2017,tanvir2017,troja2017,utsumi2017,villar2017}. }\label{fig2}
\end{figure}

The comprehensive observations of frequency-dependent lightcurves and time-dependent spectra for AT2017gfo indicate that they cannot be explained by only one single radioactivity-powered component. At least two emission components \citep{cowperthwaite2017,kasen2017,perego2017,tanaka2017,villar2017,kawaguchi2018,wanajo2018,wumr2019} or a model invoking energy injection \citep{ai2018,yu2018,li2018,ren2019} are required to account for the data of AT2017gfo. Up to now, in addition to AT2017gfo that showed relatively complete observations of kilonova properties, no kilonova candidate was found in O3\footnote{\cite{kasliwal2020} and \cite{mohite2021} combined the results of GW detections and the non-detections of EM counterparts in O3 to constrain the kilonova luminosity function.} and other potential kilonova candidates were detected in superposition with decaying sGRB afterglows. 

It is predicted that there should exist  kilonova emission diversity due to different mass ratios \citep[e.g.,][]{dietrich2017}, different kinds of merger remnant \citep[e.g.,][]{kasen2015,fujibayashi201,gill2019,kawaguchi2020,kawaguchi2021} and potential energy injections into the ejecta \citep{yu2013,metzger2014,ma2018}. The statistical analysis of kilonova emission in sGRB afterglows revealed that the kilonova brightness may have a wide range of distribution and hence confirmed that kilonova emission should be diverse \citep{gompertz2018,ascenzi2019,rossi2020,rastinejad2021}. However, on one hand,  kilonova emission is predicted to be highly viewing-angle-dependent \citep[e.g.,][]{kasen2015,martin2015,wollaeger2018,wollaeger2021,barbieri2019,bulla2019,darbha2020,zhu2020,korobkin2021} so that one can only observe these kilonova candidates in sGRB afterglows when the line of sight is close to the jet axis. On the other hand, due to the scarce and ambiguous observational data, it is hard to use these limited data to model lightcurves of cosmological kilonovae\footnote{Furthermore, dust echoes from some sGRBs occurring in star-forming galaxies may be much brighter than kilonova so that they would also affect the observations of kilonova candidates from the on-axis view \citep{lu2021}.}. We thus simply assume that all kilonovae are GW170817-like, and adopt the model with a total ejecta mass $M_{\rm ej} = 0.04\,M_\odot$ and a half-opening angle of the lanthanide-rich component $\Phi = 60^\circ$ (i.e., \texttt{nph1.0e+06\_mej0.040\_phi60}) generated by the grid of \texttt{POSSIS} \citep{bulla2019,coughlin2020} as our standard model \footnote{{We notice that \cite{dietrich2020} recently considered another kind of wind ejecta component based on the model of \texttt{POSSIS}.}}. The total ejecta mass of the lightcurve model corresponds to the constraints by AT2017gfo \citep[e.g.,][]{cowperthwaite2017,kasen2017,kasliwal2017,murguiaberthier2017,perego2017,tanaka2017,villar2017}. This viewing-angle-dependent model considers two components with different opacities, including a polar-dominated lanthanide-free component and an equatorial-dominated lanthanide-rich component. 

By setting $\theta_{\rm v} = 30^\circ$ which is relevant to the configuration of GW170817, the kilonova model we adopted can basically explain the data of AT2017gfo as shown in Figure \ref{fig2}. Only in some infrared bands, the predicted lightcurves are slightly lower than the observational data. 

\section{Viewing-angle-dependent Properties and Luminosity Function of Kilonovae and Optical Afterglows} \label{sec:3}

In Figure \ref{fig1} {and Figure \ref{fig3}}, we show the multiwavelength lightcurves of GRB170817A in the rest frame and its predicted on-axis afterglow lightcurves. Physical parameters of GRB170817A are directly adopted by the best fitting result from \cite{troja2020}. Comparing the on-axis multiwavelength lightcurves of GRB170817A with other cosmological sGRB afterglows, the physical properties of the X-ray and optical afterglow of GRB170817A are typical for sGRB afterglows. This result was also confirmed by \cite{salafia2019} and \cite{wu2019}, recently. For an on-axis configuration, AT2017gfo is fainter than the predicted associated afterglow and also fainter than most of cosmological sGRB afterglows. If GW170817 is an on-axis event, the kilonova emission would emerge as a fast-evolving bump in the bright afterglow emission at $\sim1\,{\rm day}$ after the merger. The brightness of afterglow should be more diverse than that of kilonova in the universe. The detectability of a kilonova would be significantly affected by the brightness of the associated afterglow. We generate a cosmological BNS population, investigate under what circumstances one can observe a clear kilonova signal without the pollution from an associated afterglow and predict model-dependent magnitude distributions (luminosity functions) of BNS kilonovae and afterglows. 

\subsection{Viewing-angle-dependent Properties \label{sec:3.1}}

\begin{figure*}[htbp]
    \centering
	\includegraphics[width = 0.49\linewidth , trim = 65 10 70 5, clip]{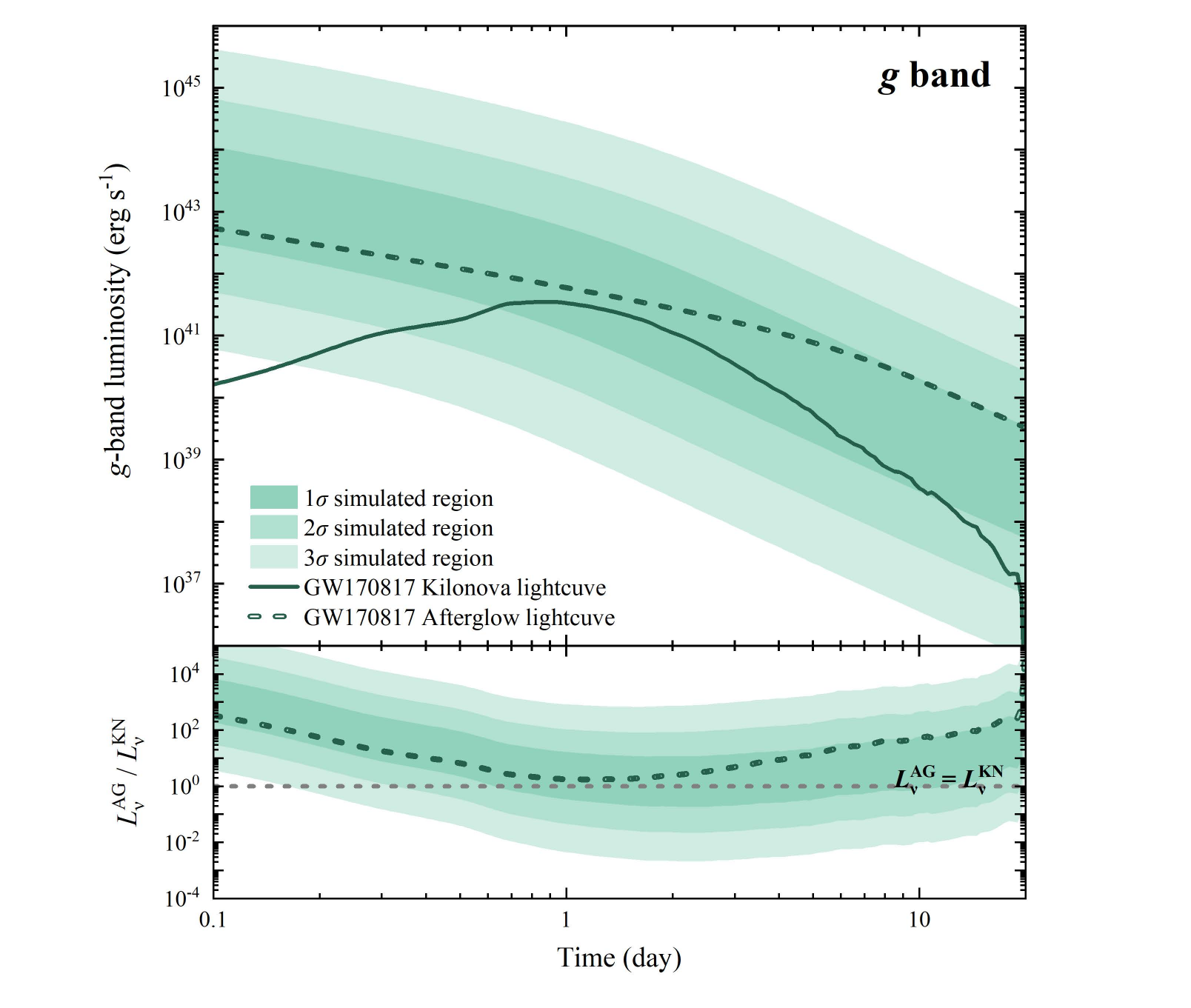}
	\includegraphics[width = 0.49\linewidth , trim = 65 10 70 5, clip]{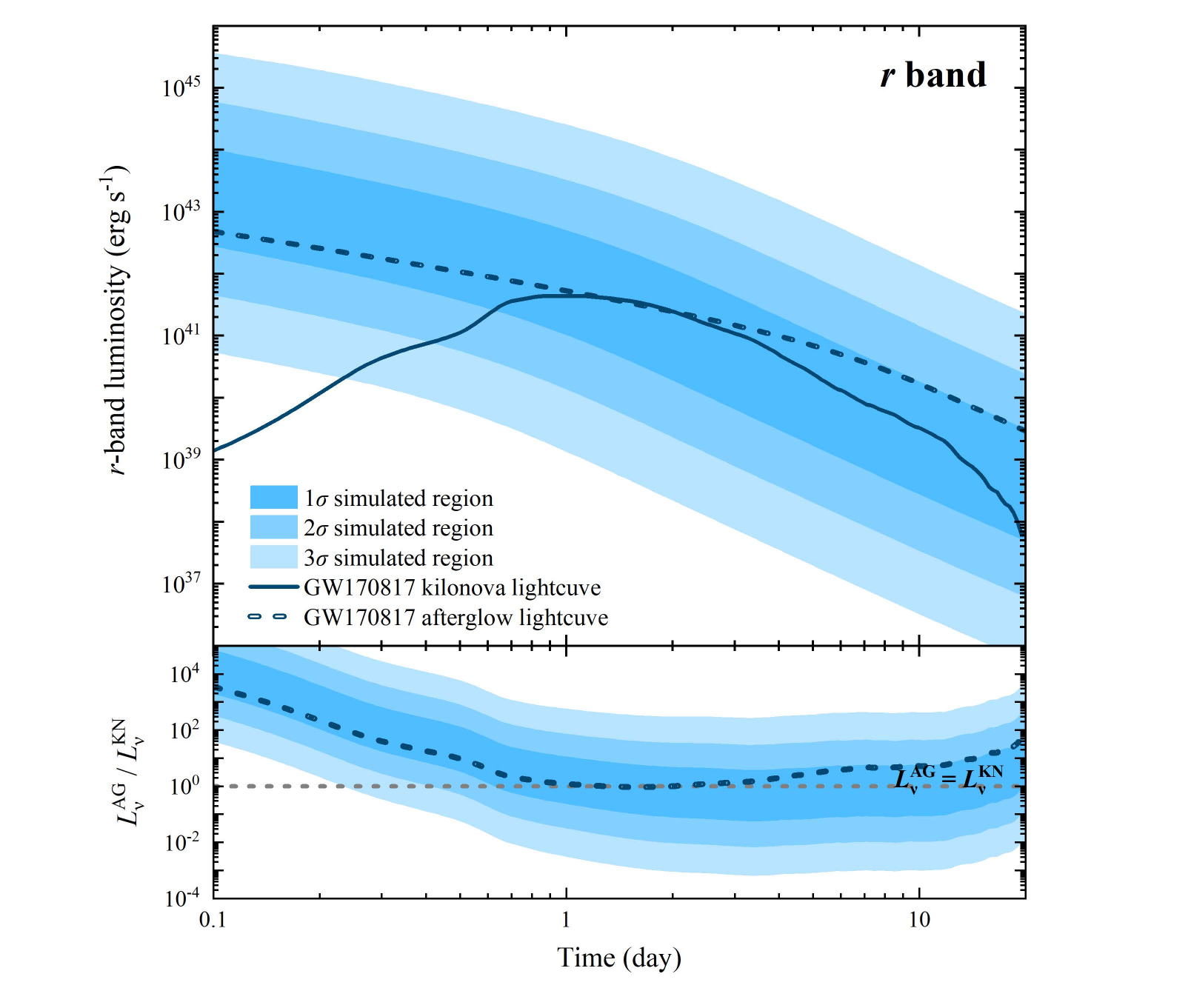}
	\includegraphics[width = 0.49\linewidth , trim = 65 20 70 5, clip]{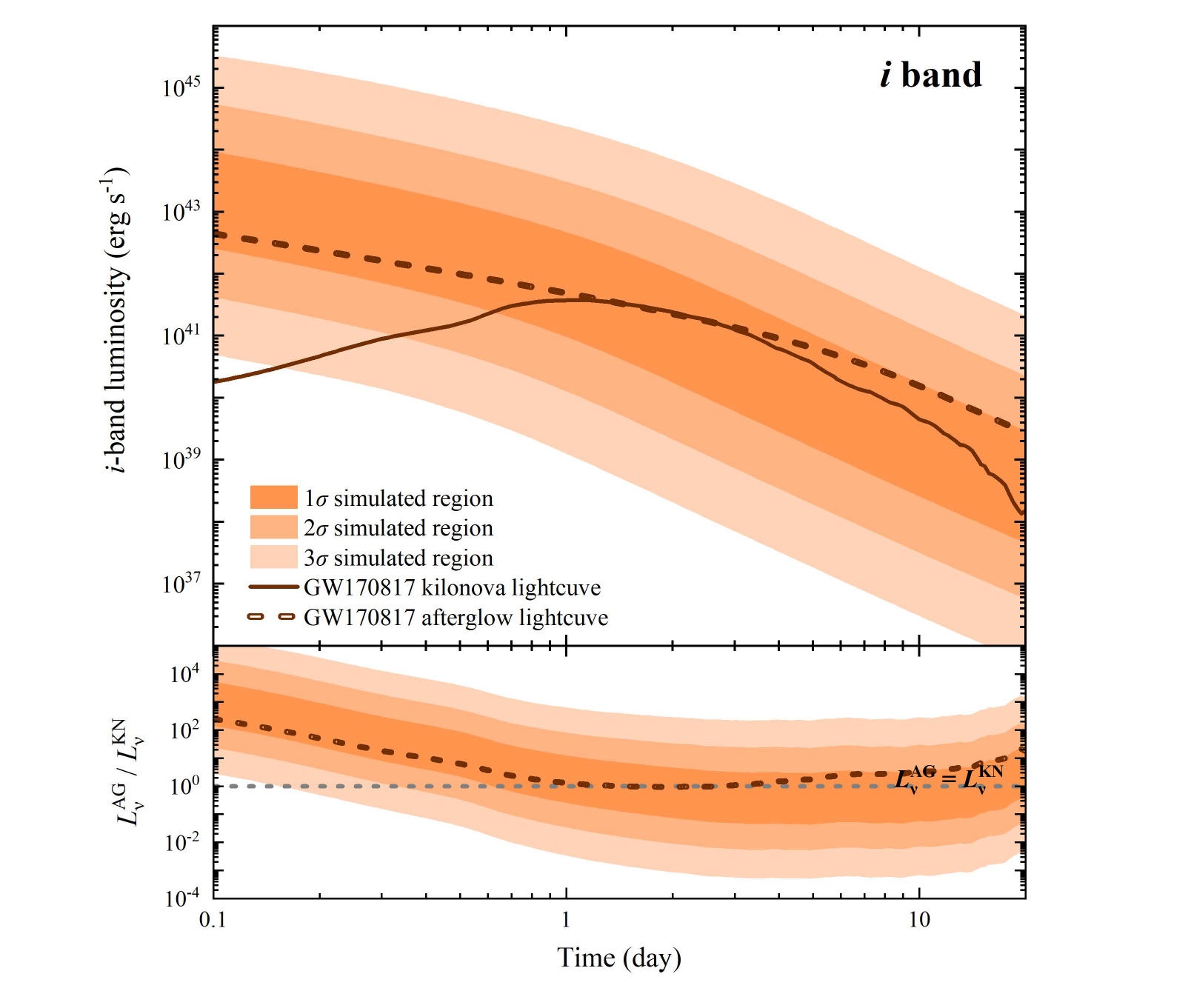}
	\includegraphics[width = 0.49\linewidth , trim = 65 20 70 5, clip]{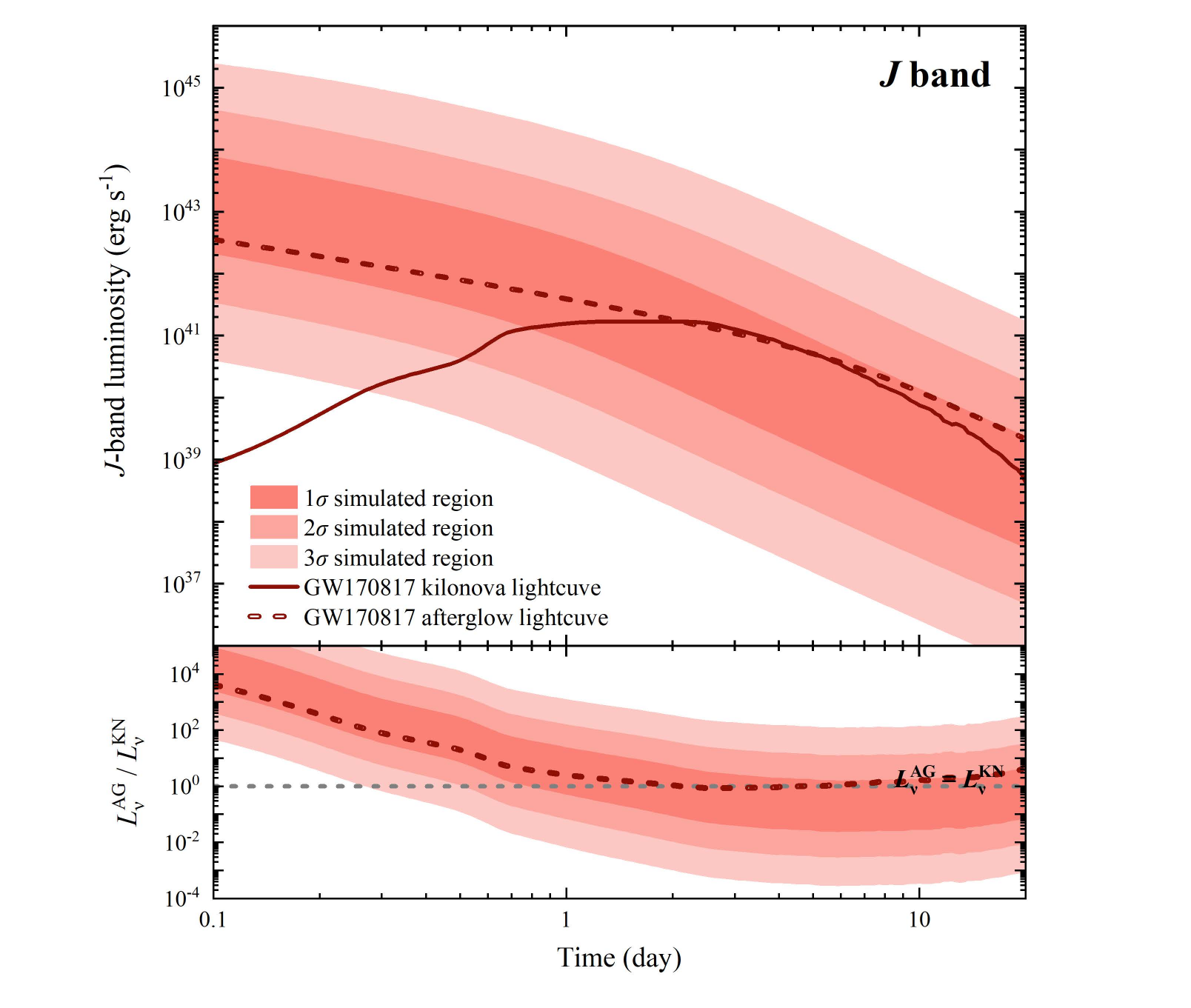}
    \caption{Kilonova and afterglow lightcurves along the polar axis (i.e., $\sin\theta_{\rm v} = 0$). The shaded regions from light to dark represent $1\sigma$, $2\sigma$ and $3\sigma$ simulated regions for cosmological afterglow. {The colored solid and dashed lines show the lightcurves of GW170817-like kilonova and afterglow, respectively.} Different panels correspond to different bands, including $griJ$ bands. The bottom sub-panels of each panel show the brightness ratio between afterglow and kilonova emissions. {The colored dashed lines are the brightness ratio between GW170817-like afterglow and kilonova along the line of sight.} The gray dashed lines show where the brightness of afterglow emission is equal to that of the  kilonova emission.}\label{fig3}
\end{figure*}

The $gri$ bands are typically used by present and future survey projects \citep[e.g., ZTF, the Multichannel Photometric Survey Telescope abbreviated as Mephisto,  the Wide Field Survey Telescope abbreviated as WFST, and the Large Synoptic Survey Telescope abbreviated as LSST\footnote{{LSST has been newly named as the Vera Rubin Observatory.}};][Er et al. 2022, in preparation; Kong et al. 2022, in preparation]{graham2019,lsst2009} while some survey instruments \citep[e.g., the Wide Field Infrared Survey Telescope abbreviated as WFIRST\footnote{{WFIRST has been newly named the Nancy Grace Roman Space Telescope.}} and the Wide Field Infrared Transient Explorer;][]{hounsell2018,lourie2020,frostig2021} designed for dedicated follow-up observations in the near-infrared bands (e.g., $J$ and $H$ bands) will operate in the near future. As shown in Figure \ref{fig3}, we plot the on-axis kilonova and afterglow lightcurves in $griJ$ bands.The luminosity ratios between afterglow and kilonova are shown in the bottom of each panel of Figure \ref{fig3}. We find that for the on-axis case, only {$\lesssim50\%$} afterglows are dimmer than the associated kilonovae at the kilonova peak time, in which case the kilonovae could be identified as a fast-evolving significant bump on top of the afterglow lightcurve. {More specifically, $\sim15\%$ of the kilonovae would be more than tenfold as bright as the associated afterglows, while $\sim35\%$ of the on-axis kilonovae would be brighter than the afterglows by a factor of $\sim1-10$.} The brightness of {$\gtrsim50\%$} of the afterglows is always larger than that of the associated kilonovae, especially during the early and late stages. In such a case, kilonova emission would be outshone by the afterglow emission and is difficult to be identified. {If one wants to identify the potential kilonova emission in afterglow emission via optical serendipitous observations \citep[e.g.,][]{berger2013,jin2015,wu2021,ahumada2021}, a few criteria are required: 1. the event is nearby so that the observed emission can last for $\gtrsim1 - 2\,{\rm days}$; 2. the early-stage afterglow lightcurve has a high-cadence record; 3. a significant excess of kilonova emission with respect to the afterglow lightcurve.} The best detection period would be a few days after the BNS mergers, when the afterglow-to-kilonova luminosity ratio reaches the minimum value. Meanwhile, we find that the infrared bands, e.g., $i$ and $J$, are better for searching and identifying kilonovae from afterglow lightcurves, thanks to the longer durations for kilonova emission in redder bands.

\begin{figure*}[htpb]
    \centering
	\includegraphics[width = 0.32\linewidth , trim = 65 10 70 5, clip]{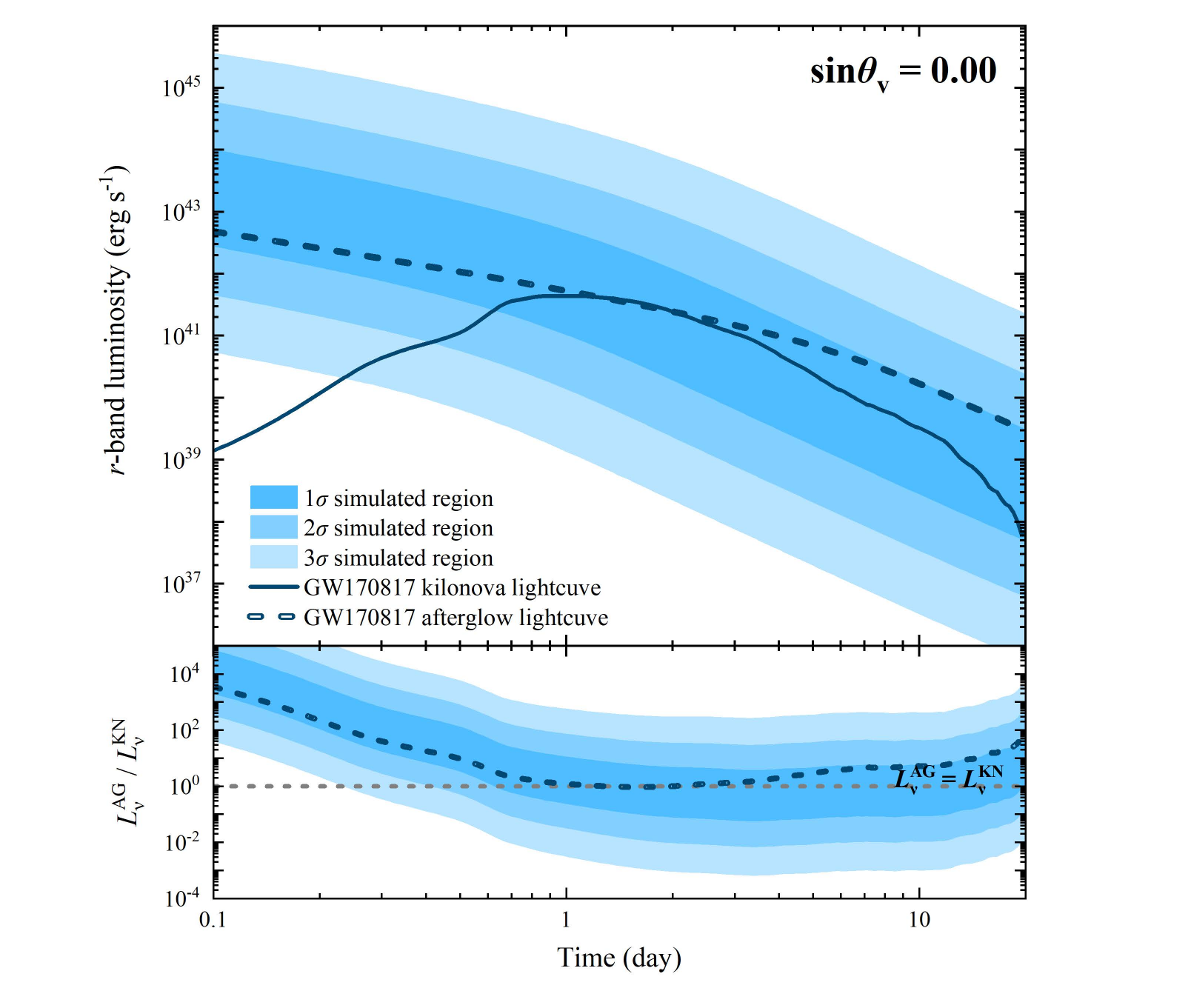}
	\includegraphics[width = 0.32\linewidth , trim = 65 10 70 5, clip]{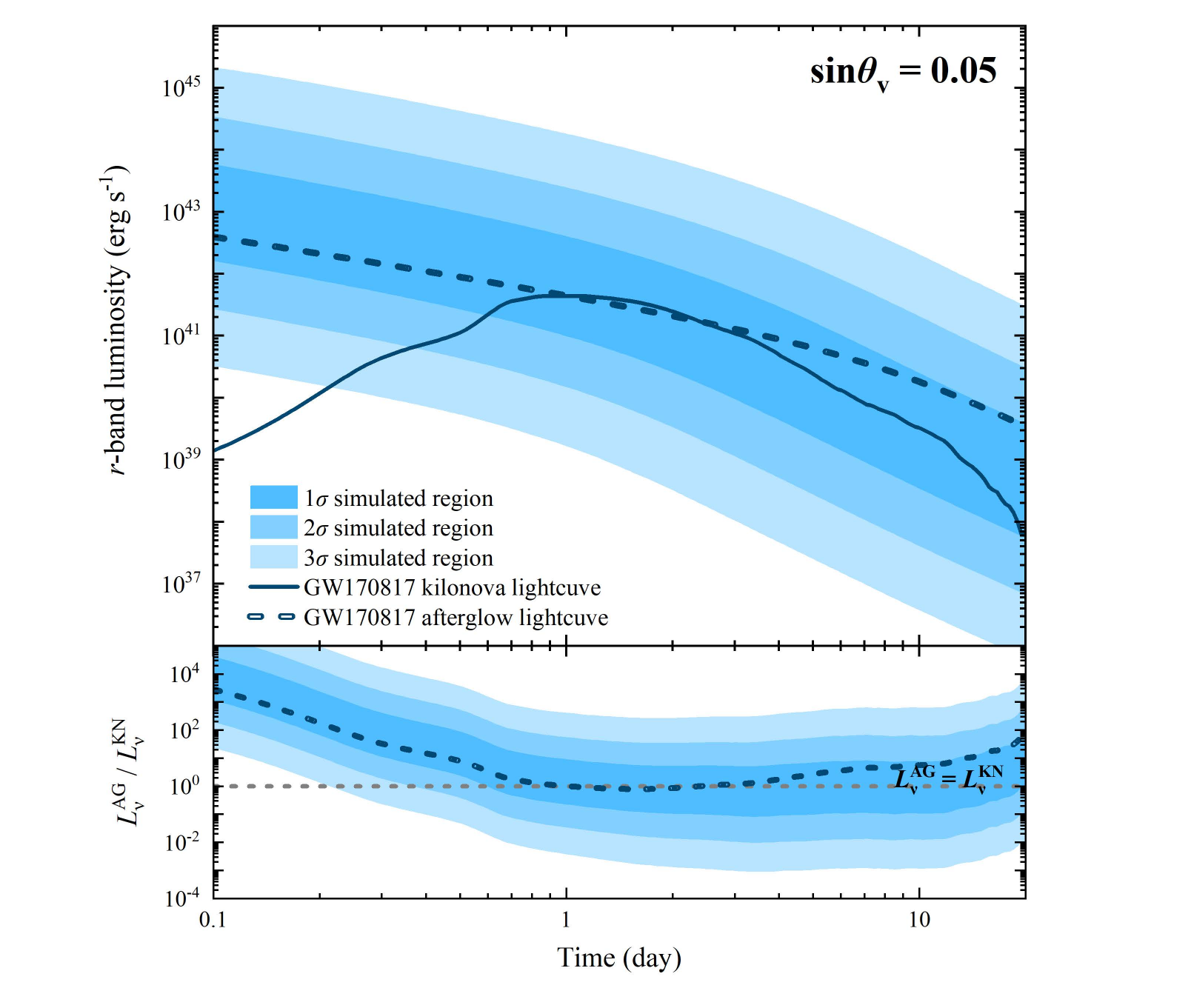}
	\includegraphics[width = 0.32\linewidth , trim = 65 10 70 5, clip]{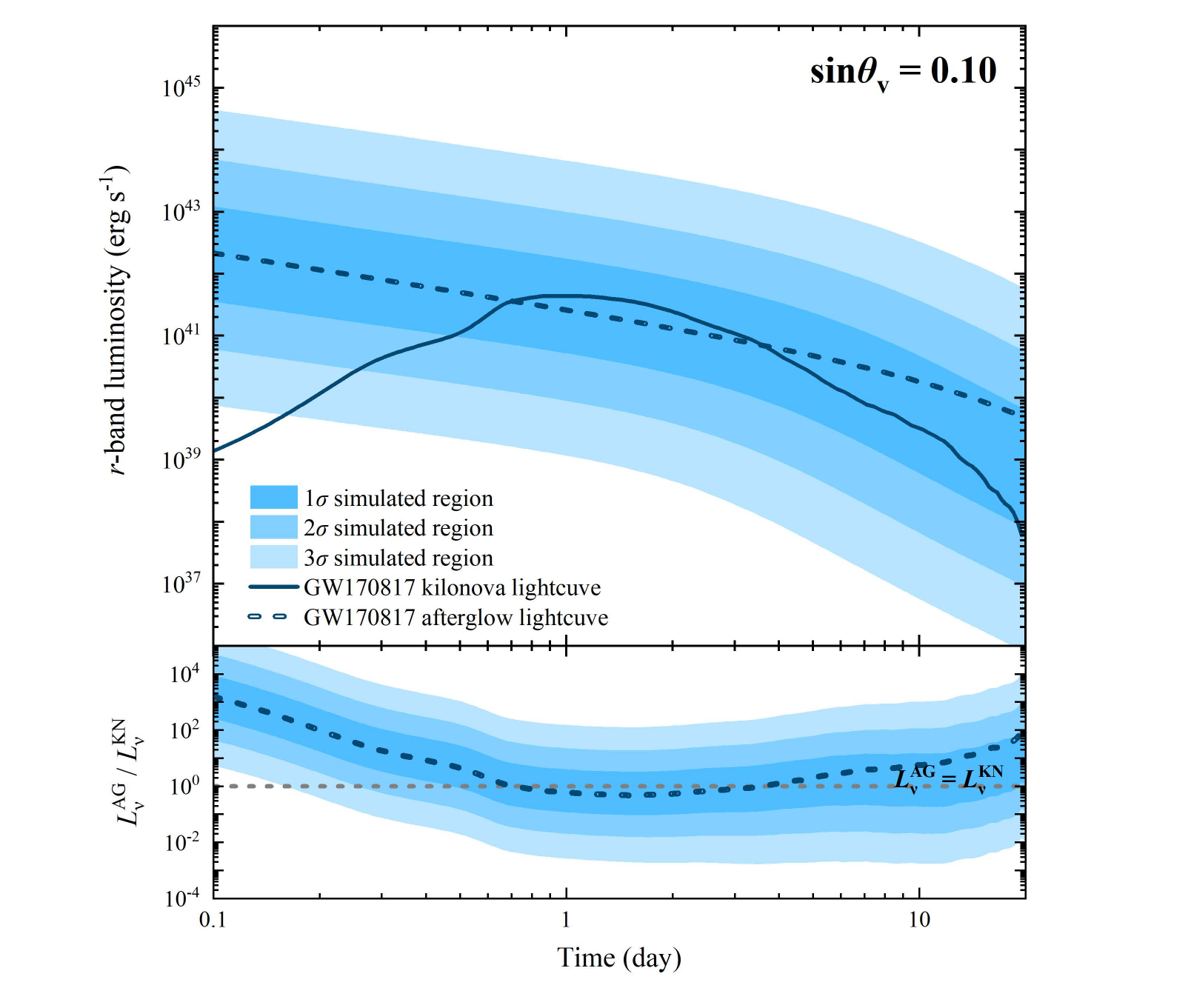}
	\includegraphics[width = 0.32\linewidth , trim = 65 20 70 5, clip]{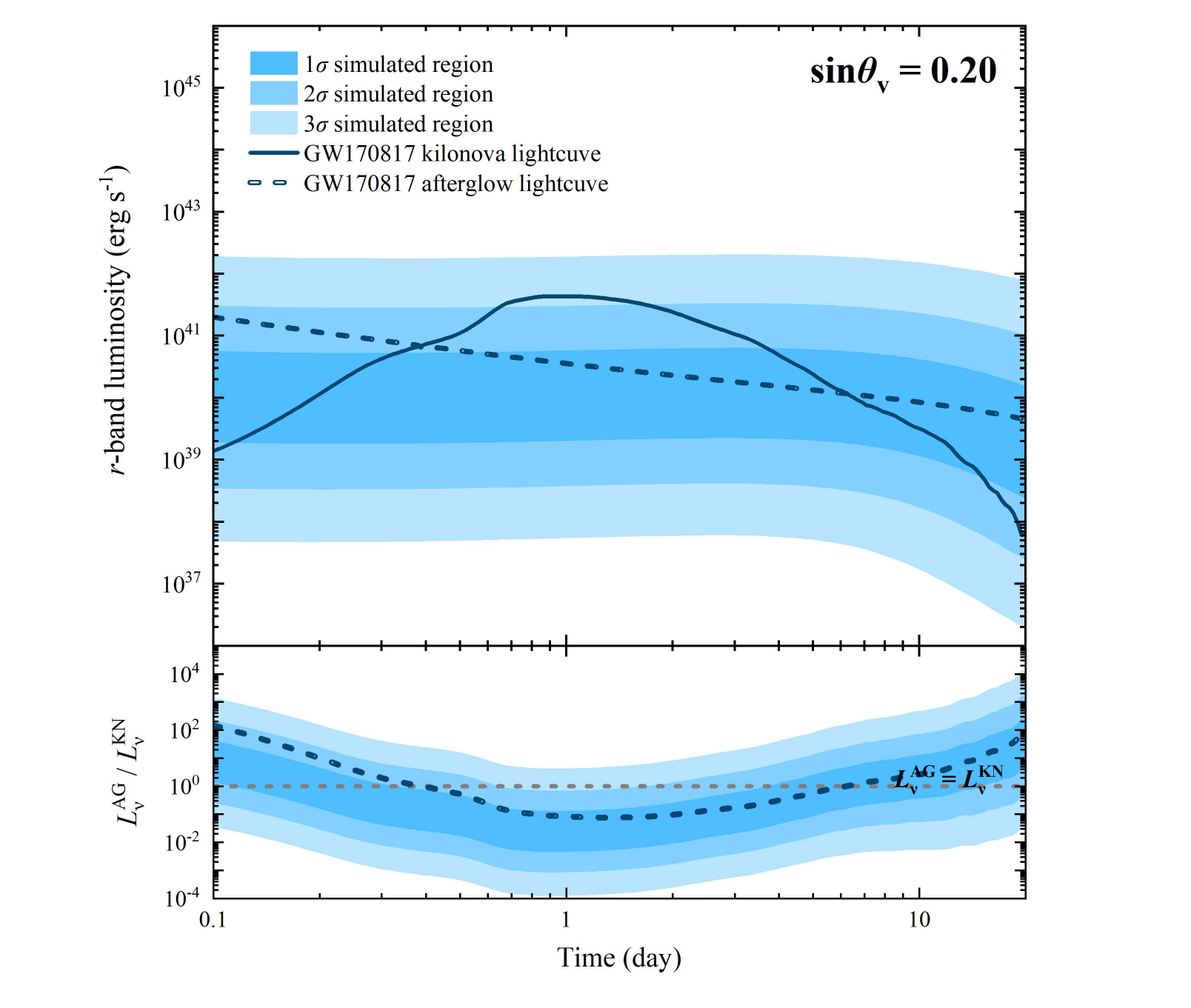}
	\includegraphics[width = 0.32\linewidth , trim = 65 20 70 5, clip]{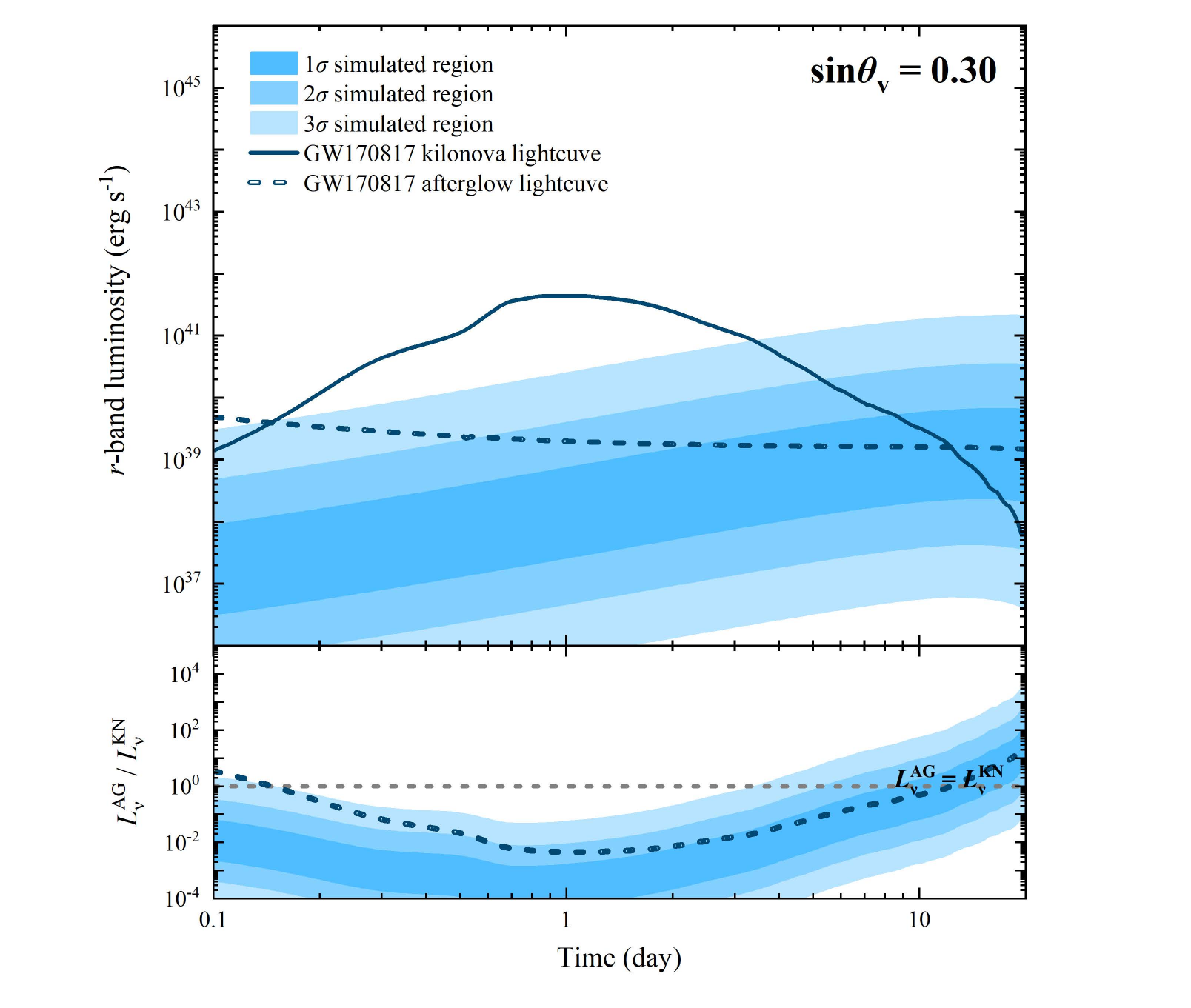}
	\includegraphics[width = 0.32\linewidth , trim = 65 20 70 5, clip]{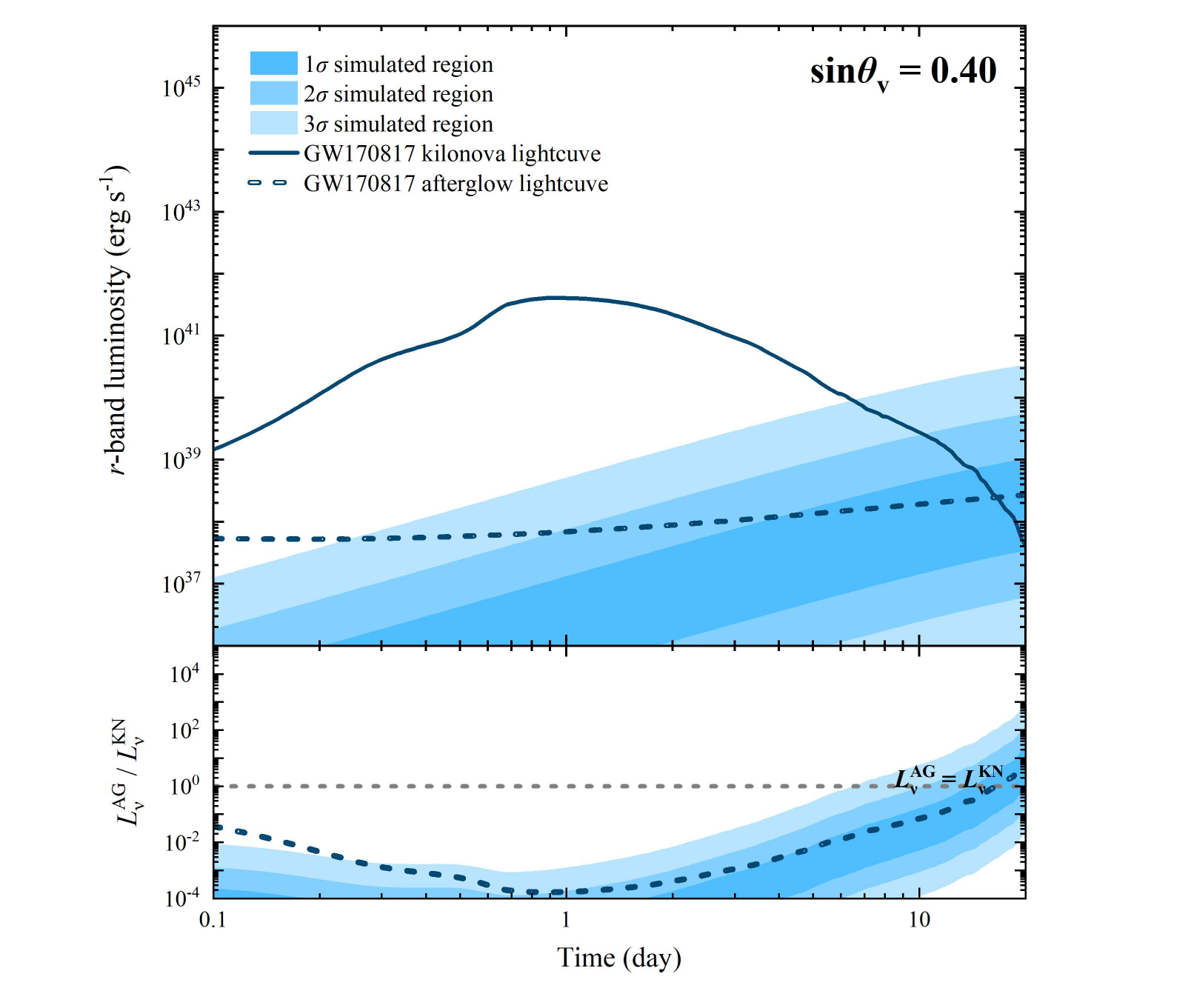}
    \caption{Similar to Figure \ref{fig3}, but for the $r$-band afterglow lightcurves with the consideration of several different viewing angles, including $\sin\theta_{\rm v} = 0,\,0.05,\,0.1,\,0.2,\,0.3,\,{\rm and}\,0.4$. }\label{fig4}
\end{figure*}

Since the afterglow radiation is significantly beamed due to the relativistic outflow, the afterglow lightcurve would highly depend on the viewing direction. Figure \ref{fig4} shows the kilonova and afterglow lightcurves with different viewing directions in the $r$ band. One can conclude that the larger the viewing angle, the fainter the afterglow emission and the more significant the kilonova contribution. For the afterglow emission with $\sin\theta_{\rm v} \lesssim 0.10$ ($\theta_{\rm v}\lesssim2\theta_{\rm c}$), the EM signals from most BNS mergers would still be afterglow-dominated. {The observations of kilonovae with $\sin\theta_{\rm v} \lesssim 0.20$ may easily be polluted by the associated afterglows, because observable kilonovae would always emerge as a fast-evolving bump on top of the afterglow emissions.} When $\sin\theta_{\rm v}\gtrsim0.20$, the peak luminosity of kilonova emission from most BNS mergers is always much brighter than the afterglow luminosity at $\sim1\,{\rm day}$ after the merger. In this case, it is only when the kilonova becomes dim, i.e., after $\sim5-10\,{\rm day}$, that one can observe the off-axis jet afterglow.

\subsection{Luminosity Function}

\begin{figure*}[htbp]
    \centering
	\includegraphics[width = 0.32\linewidth , trim = 80 95 100 25, clip]{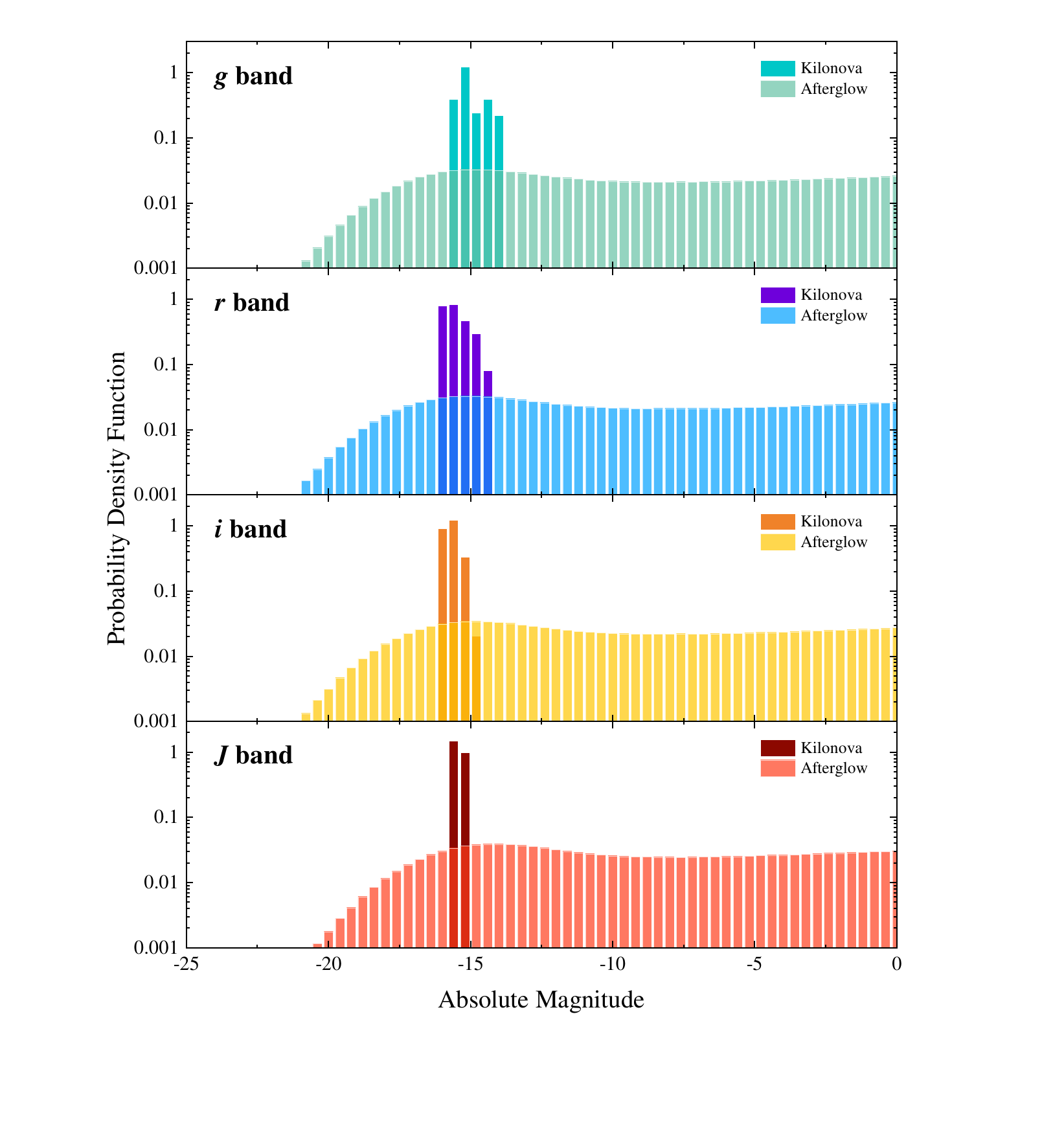}
	\includegraphics[width = 0.32\linewidth , trim = 80 95 100 25, clip]{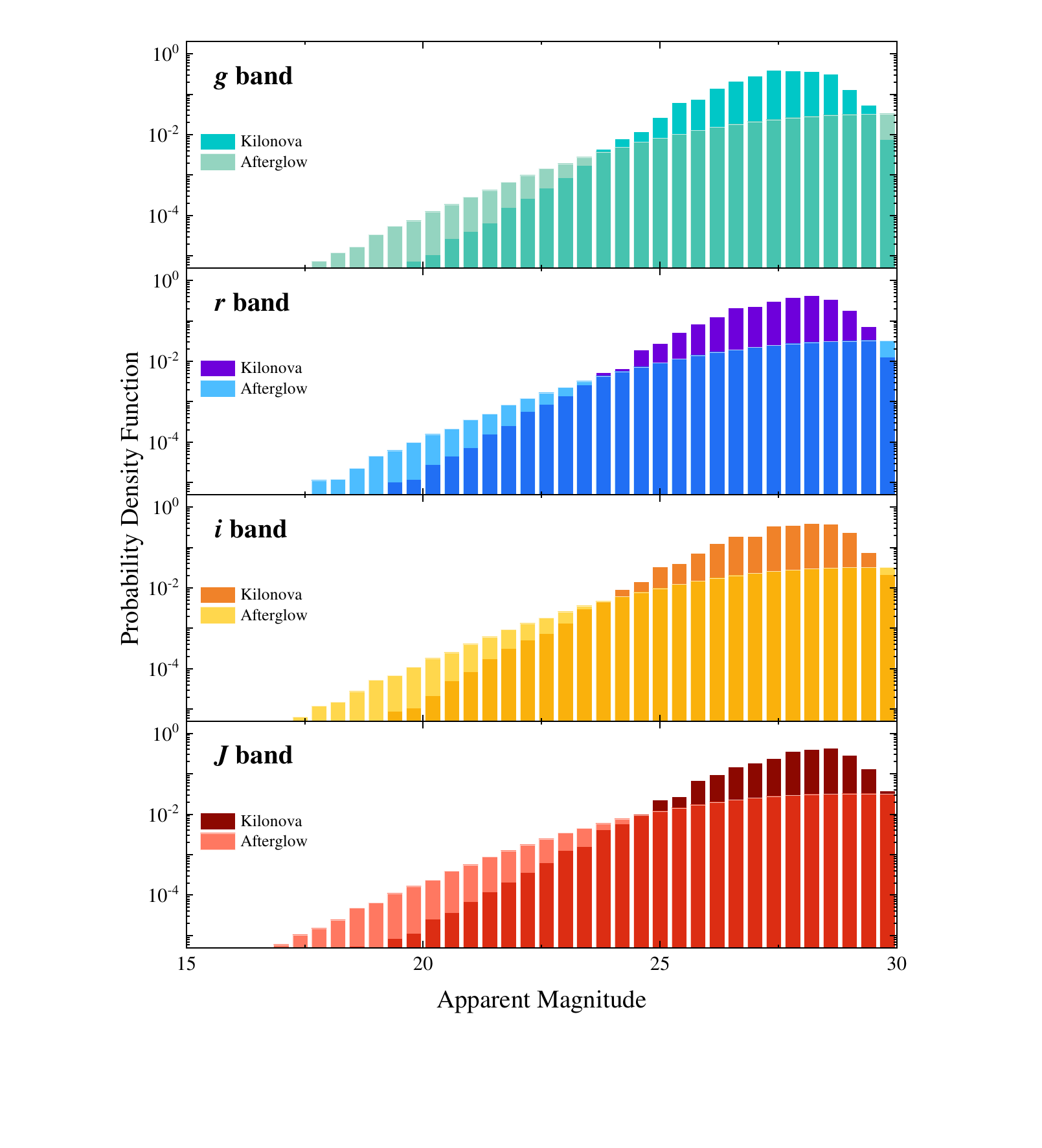}
	\includegraphics[width = 0.32\linewidth , trim = 80 95 100 25, clip]{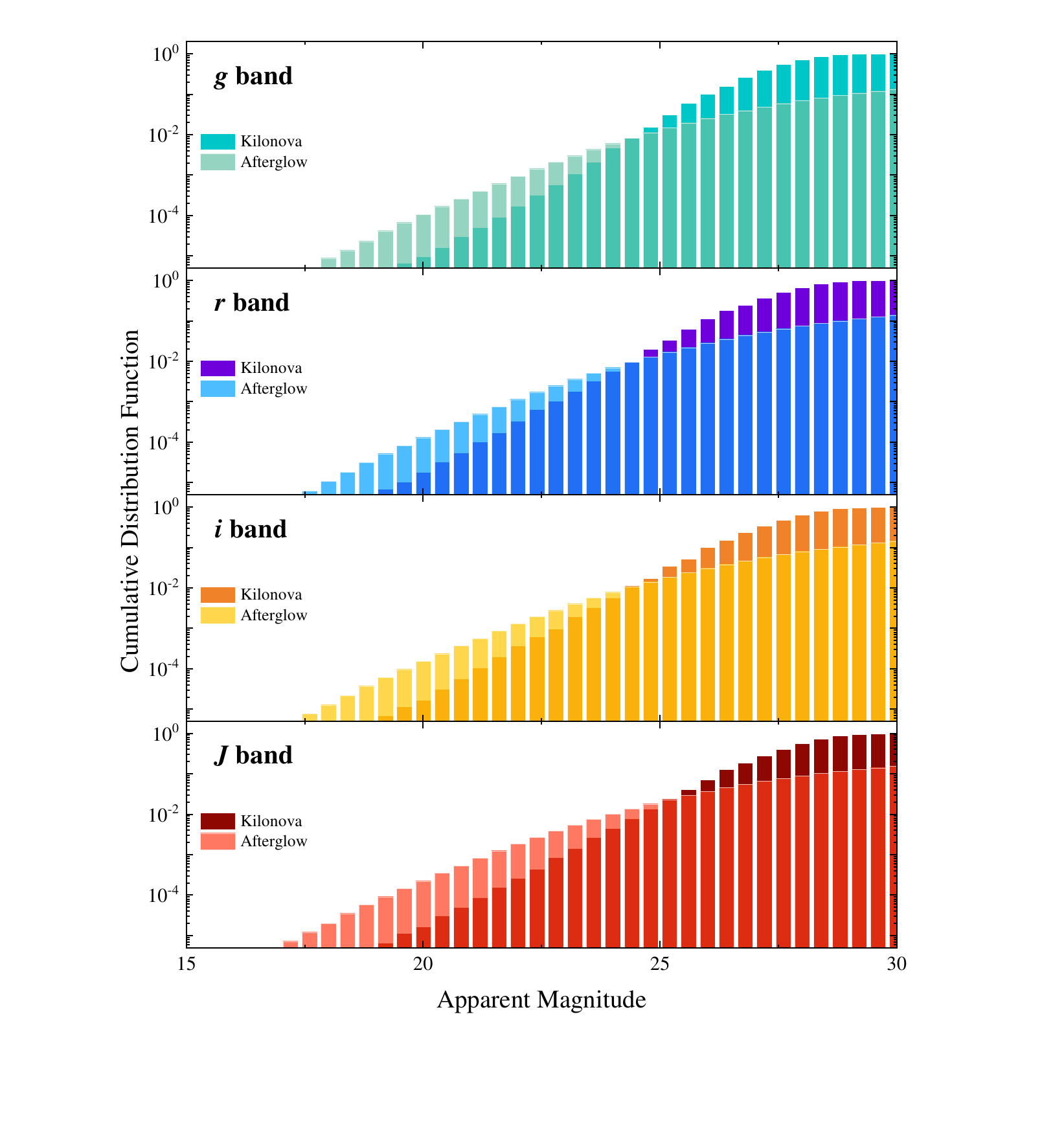}
    \caption{Panels from left to right represent the PDFs of the absolute magnitude, the PDFs of the apparent magnitude, and the cumulative distributions of the apparent magnitude for kilonova and afterglow emissions in the $griJ$ bands, respectively. Here, the bin width of the histograms is set as $\Delta = 0.4$ mag.}\label{fig5}
\end{figure*}

{With the redshift distribution $f(z)$, we randomly simulate a group of $n_{\rm sim} = 5\times 10^6$ BNS events in the universe based on Equation (\ref{equ:merger_rate}). The distribution of viewing angle $\theta_{\rm v}$ is adopted to be uniform. We simulate the viewing-angle-dependent kilonova and afterglow emissions for each simulated BNS event based on the kilonova and afterglow models described in the previous section. We emphasize that the magnitudes of the kilonova and afterglow emissions represent emergent magnitudes obtained in the Earth frame in the following discussion.}

The left panel of Figure \ref{fig5} shows the probability density functions (PDFs) of the kilonova peak absolute magnitude and the absolute magnitude for the associated afterglow at the kilonova peak time in $griJ$ bands. {We notice that the following calculations and discussions for afterglows should be conservative considering that the luminosity function of afterglow is calculated at the peak time of the associated kilonova.} One can see that compared with afterglow emission, the magnitude distribution of kilonova has a much narrower distribution. The absolute magnitude distribution of kilonova is mainly peaked at $\sim-16\,{\rm mag}$. Meanwhile, the redder the observing band, the narrower the magnitude distribution and the higher the luminosity. The distribution spans $\sim3\,{\rm mag}$ in optical while it spans $\sim1-2\,{\rm mag}$ in infrared. For afterglows (including on-axis and off-axis cases), a small fraction ($\sim10\%$) of afterglows would be brighter than kilonovae {at the kilonova peak time}. As discussed in Section \ref{sec:3.1}, these EM signals from BNS mergers should be afterglow-dominated events which are always observed near the polar direction. Since off-axis afterglows always have a much lower brightness than kilonovae, most EM signatures from BNS mergers along the line of sight should be kilonova-dominated.

In the middle and right panel of Figure \ref{fig5}, after considering the cosmological redshift distribution of the BNS mergers, we respectively show the PDFs and cumulative distributions for the apparent magnitudes of kilonova and afterglow in different bands. One can see that the maximum possible apparent magnitude for cosmological kilonovae is $\sim27-28\,{\rm mag}$. The distribution range of the apparent magnitude of kilonovae is still narrower than that of afterglows. For the wide-field survey projects whose search depths are relatively shallow, one can detect bright afterglows with a relatively higher probability than kilonovae. More specifically, for the search depth of $\sim20-21\,{\rm mag}$ ($\sim23-24\,{\rm mag}$), the detection probability of afterglows by survey projects is $\sim10-50$ ($\sim1-5$) times that of kilonovae. Only if the search depth reaches $\sim24-25\,{\rm mag}$ can one discover a similar amount of kilonova and afterglow events. ZTF has a search depth of $20.8\,{\rm mag}$ ($21.6\,{\rm mag}$) with an exposure time of $30\,{\rm s}$ ($300\,{\rm s}$) in the $g$ band. In 13 months of ZTF science validation, \cite{andreoni2021} reported seven independent optically-discovered GRB afterglows without any detection of kilonova candidates. Among these discovered afterglows, {two} events were inferred to be associated with an SGRB. {One of these two events likely  originated from a collapsar \citep{ahumada2021,zhang2021,rossi2021}.} These searches by ZTF were supported by our simulation results. {Mephisto, WFST, and LSST will have a search depth of 22.4\,mag (24.2\,mag), 23.0\,mag (24.2\,mag), and 25.1\,mag (26.2\,mag) with an exposure time of $30\,{\rm s}$ ($300\,{\rm s}$) in the $g$ band. It is expected that the detection rate of kilonovae could have the same order of magnitude as that of afterglows via serendipitous observations in the era of Mephisto, WFST, LSST, etc. }

We will discuss the specific detection rates of kilonova and afterglow by GW-triggered target-of-opportunity observations and serendipitous observations with different search magnitudes in \citetalias{zhu2021kilonovaafterglow} in detail. 

\section{Color evolution} \label{sec:4}

In this section, we discuss color evolution of kilonova and afterglow emissions, which would be helpful to identify these transients using future multiwavelength observations. Despite the fast evolution of kilonova and afterglow emissions, some of these transients might have been recorded in the archival data of survey projects. Color evolution analyses might help to identify them if multi-color data are available.

\begin{figure*}[htbp]
    \centering
	\includegraphics[width = 0.49\linewidth , trim = 55 30 60 25, clip]{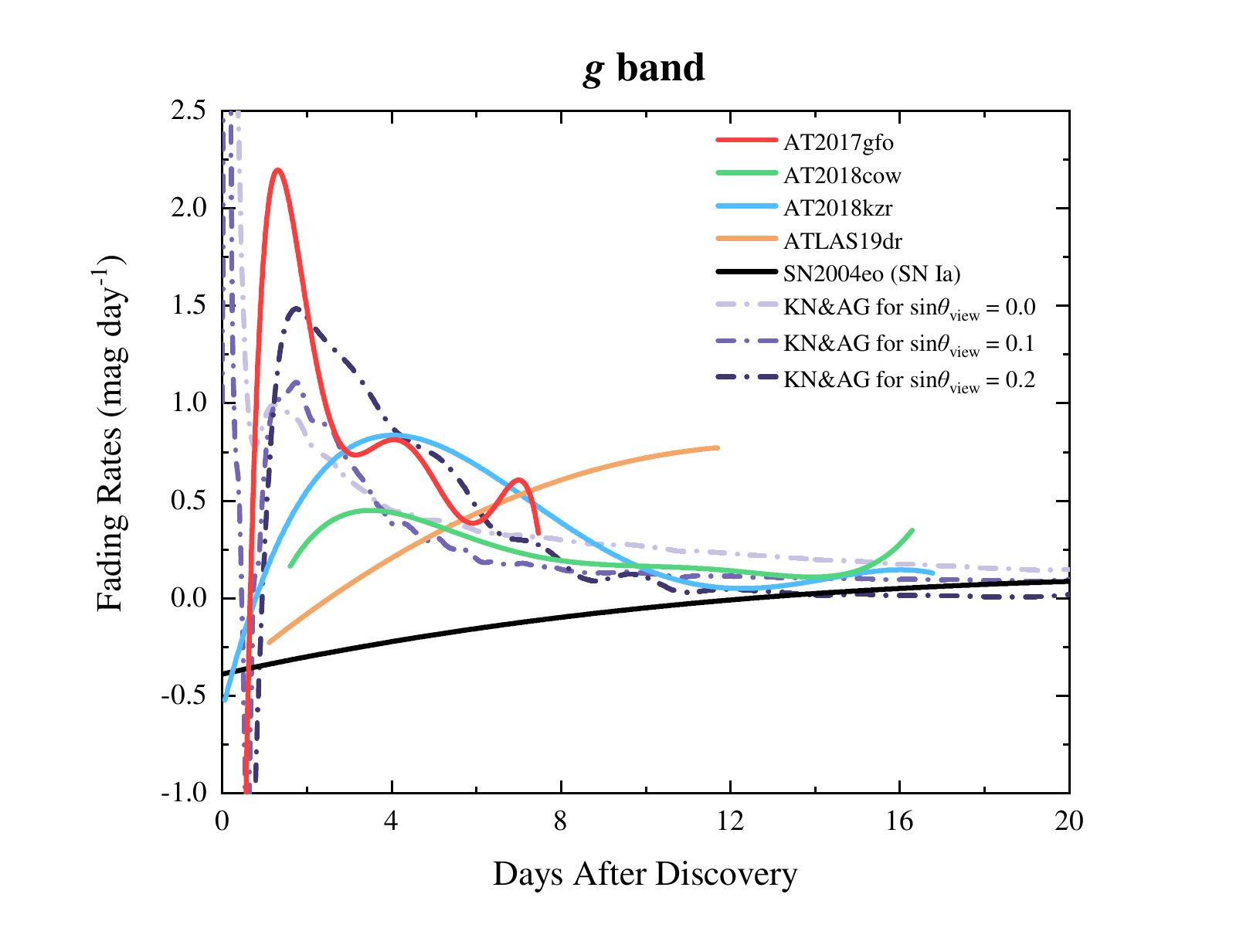}
	\includegraphics[width = 0.49\linewidth , trim = 55 30 60 25, clip]{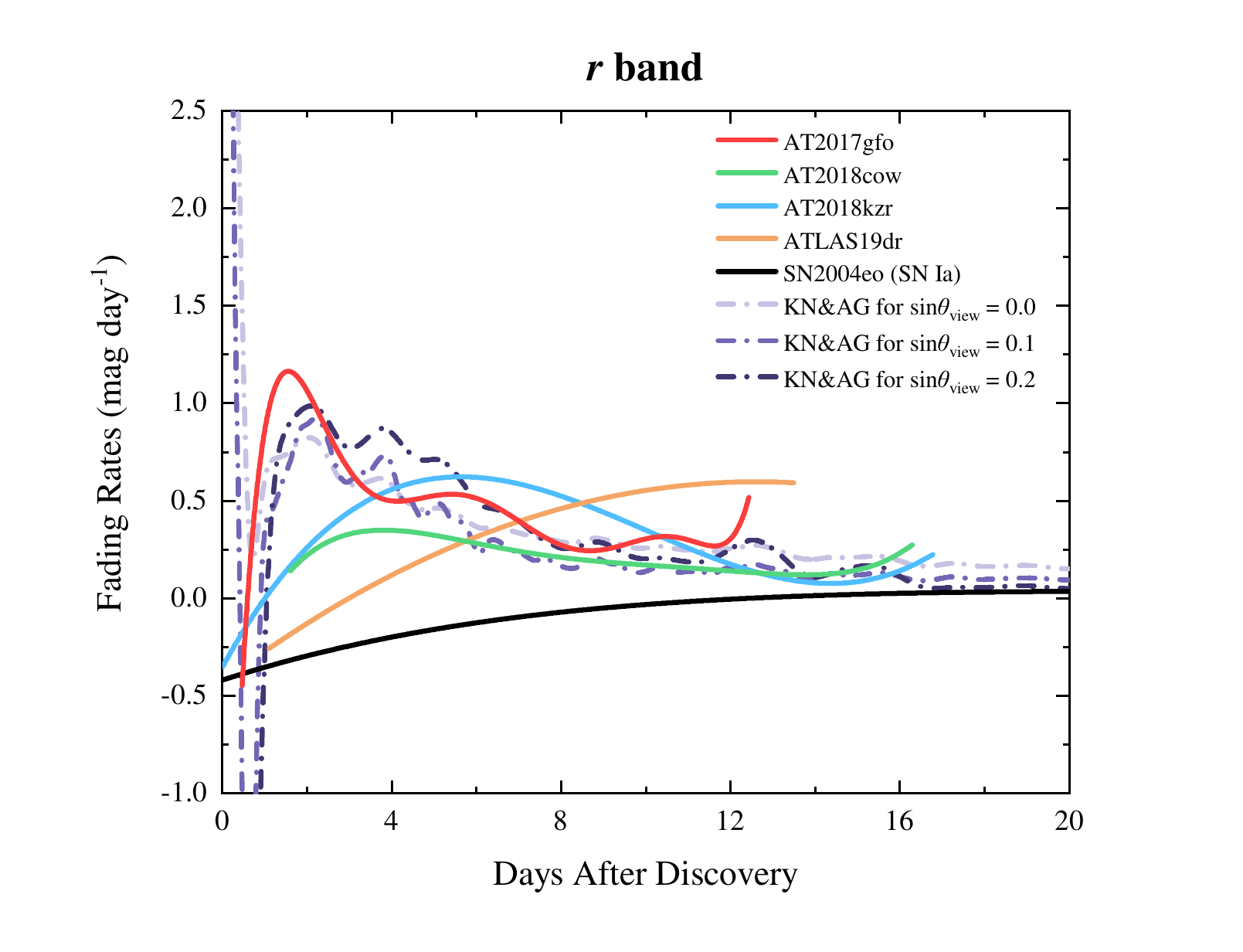}
	\includegraphics[width = 0.49\linewidth , trim = 55 30 60 25, clip]{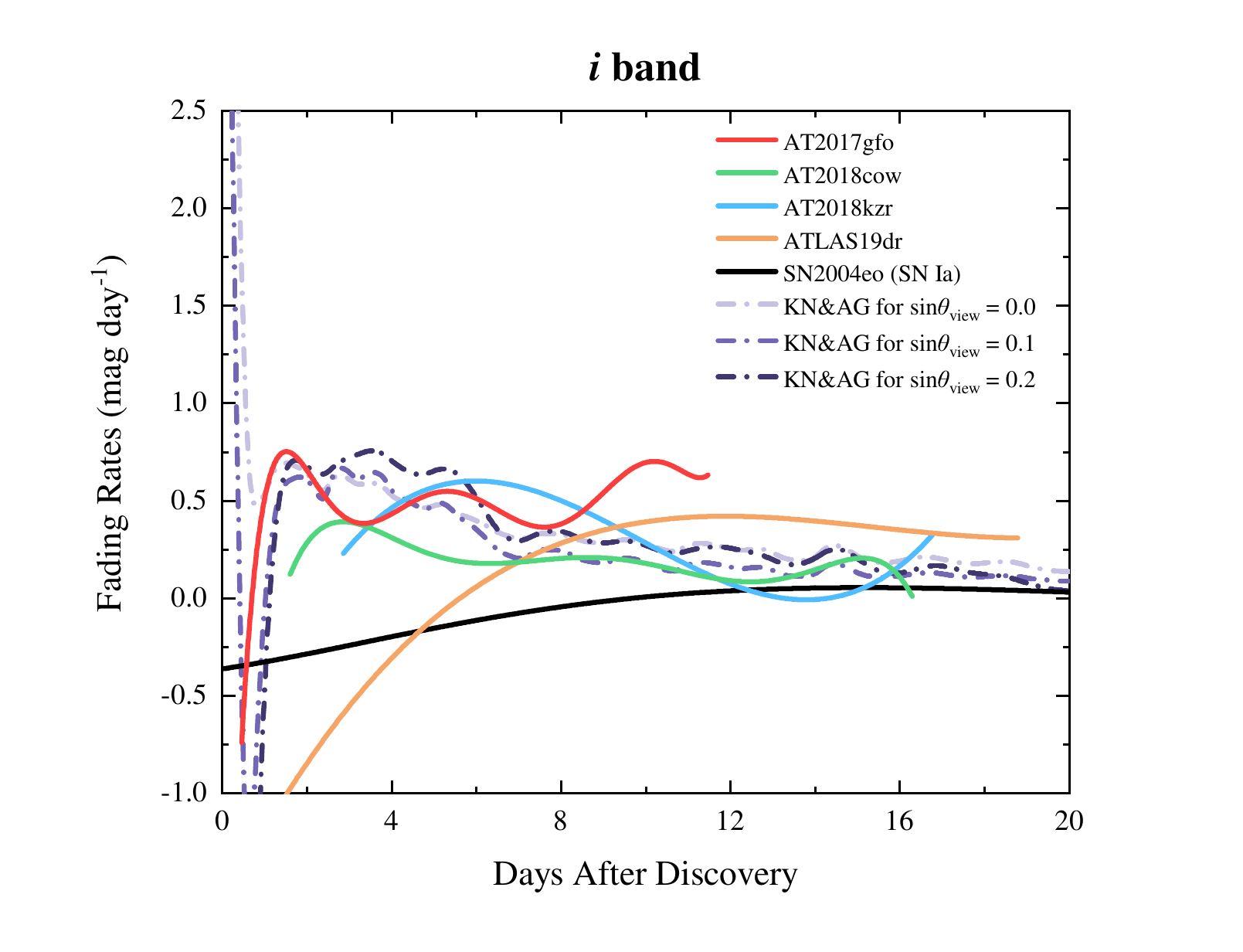}
	\includegraphics[width = 0.49\linewidth , trim = 55 30 60 25, clip]{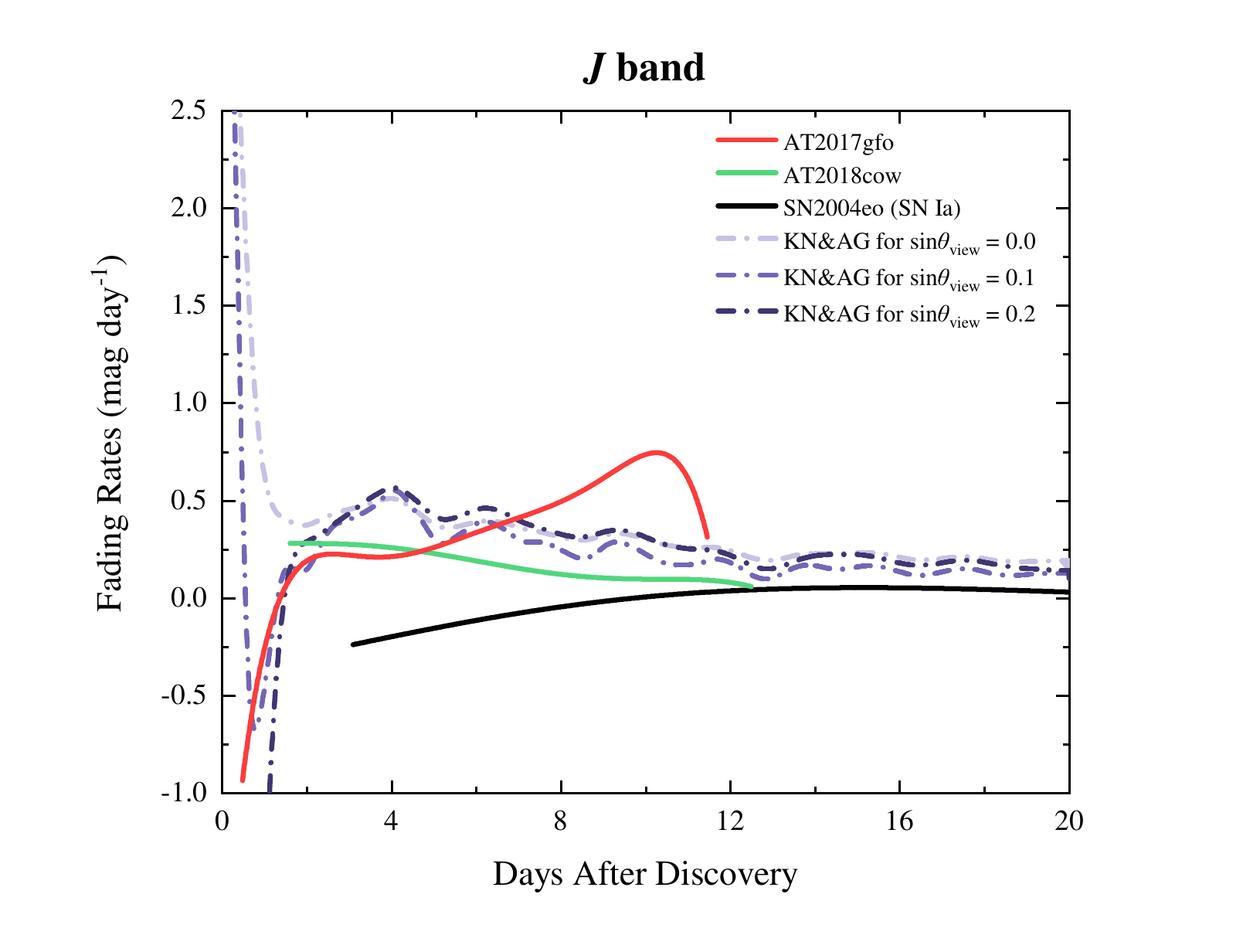}
    \caption{The $griJ$-band fading rates of AT2017gfo (red lines), AT2018cow (green lines), AT2018kzr (blue lines), ATLAS19dqr (orange lines), SN2004eo (SN Ia; black lines), and the total emission of our modeled GW170817-like kilonova and afterglow along different lines of sight (dashed-dotted lines).}\label{fig6}
\end{figure*}

In Figure \ref{fig6}, we show the $griJ$-band fading rates of AT2017gfo, the total emission of the modeled GW170817-like kilonova and afterglow along three lines of sight (i.e., $\sin\theta_{\rm v} = 0.0,\,0.1,\,{\rm and}\,0.2$), and some other well studied rapid-evolving transients, including AT2018cow \citep{prentice2018,perley2019}, AT2018kzr \citep{mcbrien2019}, ATLAS19dqr \citep{chen2020}, and SN2004eo \citep[SN Ia;][]{mazzali2008}. The fading rate of each transient is fitted by a polynomial function. One can directly use fading rates to reject supernovae which comprises the majority of the transient survey background. {A wide-range of kilonova model grids from \cite{dietrich2020} showed that the fading rates of the simulated kilonova populations are almost faster than $0.3\,{\rm mag}\,{\rm day}^{-1}$. Thus,} \cite{andreoni2021b,andreoni2021} intended to select kilonova and afterglow candidates from ZT survey database {using the pipeline of \texttt{ZTFReST} \citep[][ a ZTF software to search for kilonovae and fast-evolving transients]{andreoni2021b}} by considering recorded sources having significant fading rates faster than $0.3\,{\rm mag}\,{\rm day}^{-1}$. However, as shown in Figure \ref{fig6}, this selection condition may hardly {directly identify kilonovae from discovered fast-evolving transients in the survey dataset.} Comparing with other fast-evolving optical transients except for SN Ia, AT2017gfo and the total emission of {GW170817-like} kilonova plus afterglow have a relatively larger fading rate at the early stage ($\sim1\,{\rm day}$), especially in the optical bands. For example, {if all kilonovae were GW170817-like,} for the $g$ band, the largest fading rate of kilonovae could reach $\sim2\,{\rm mag\,day^{-1}}$ while other transients always have a fading rate $\lesssim1\,{\rm mag\,day^{-1}}$. After $\sim1-1.5\,{\rm day}$, the optical fading rate of AT2017gfo would be similar to those of AT2018cow and AT2018kzr. Thus, this requires the cadence time of survey projects to be shorter than $\sim1-2\,{\rm day}$. By only using $g$ or $r$ bands to search for kilonovae, if the survey projects miss their early-stage observations, one cannot use the fading rate to identify {GW170817-like} kilonovae from fast-evolving transients. {A fraction of kilonova populations in the universe might have relatively slower evolution at the early stage compared with AT2017gfo. We notice that distinguishing these slower-evolving kilonova populations from numerous types of fast-evolving transients by measuring single-band fading rates may still be difficult, even if the serendipitous survey projects do not miss the early-stage observations.}  Furthermore, it may be hard to use infrared fading rate to identify kilonova since all the transients have similar decay rates.

\begin{figure*}[htpb]
    \centering
	\includegraphics[width = 0.318\linewidth , trim = 72 20 90 60, clip]{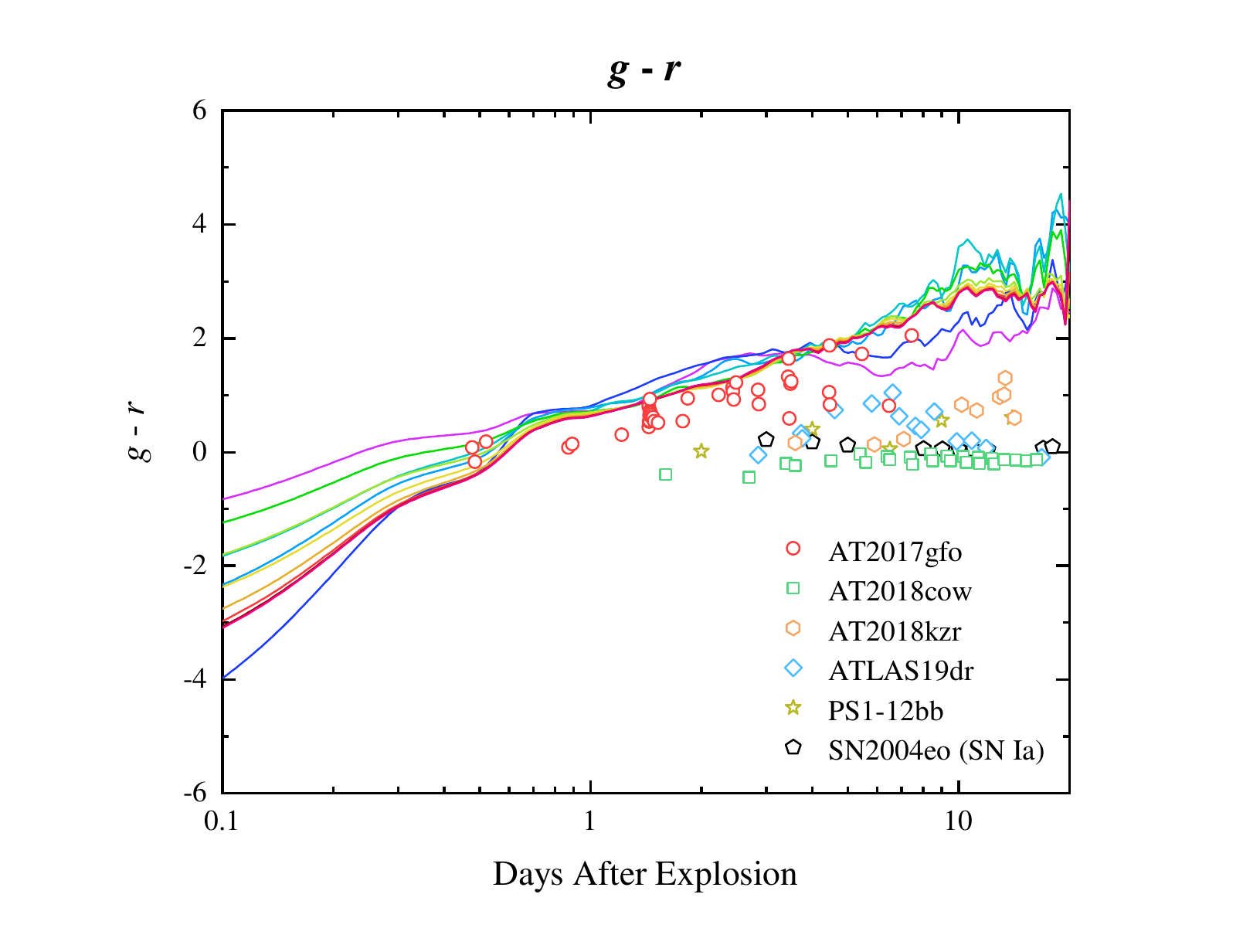}
	\includegraphics[width = 0.318\linewidth , trim = 72 20 90 60, clip]{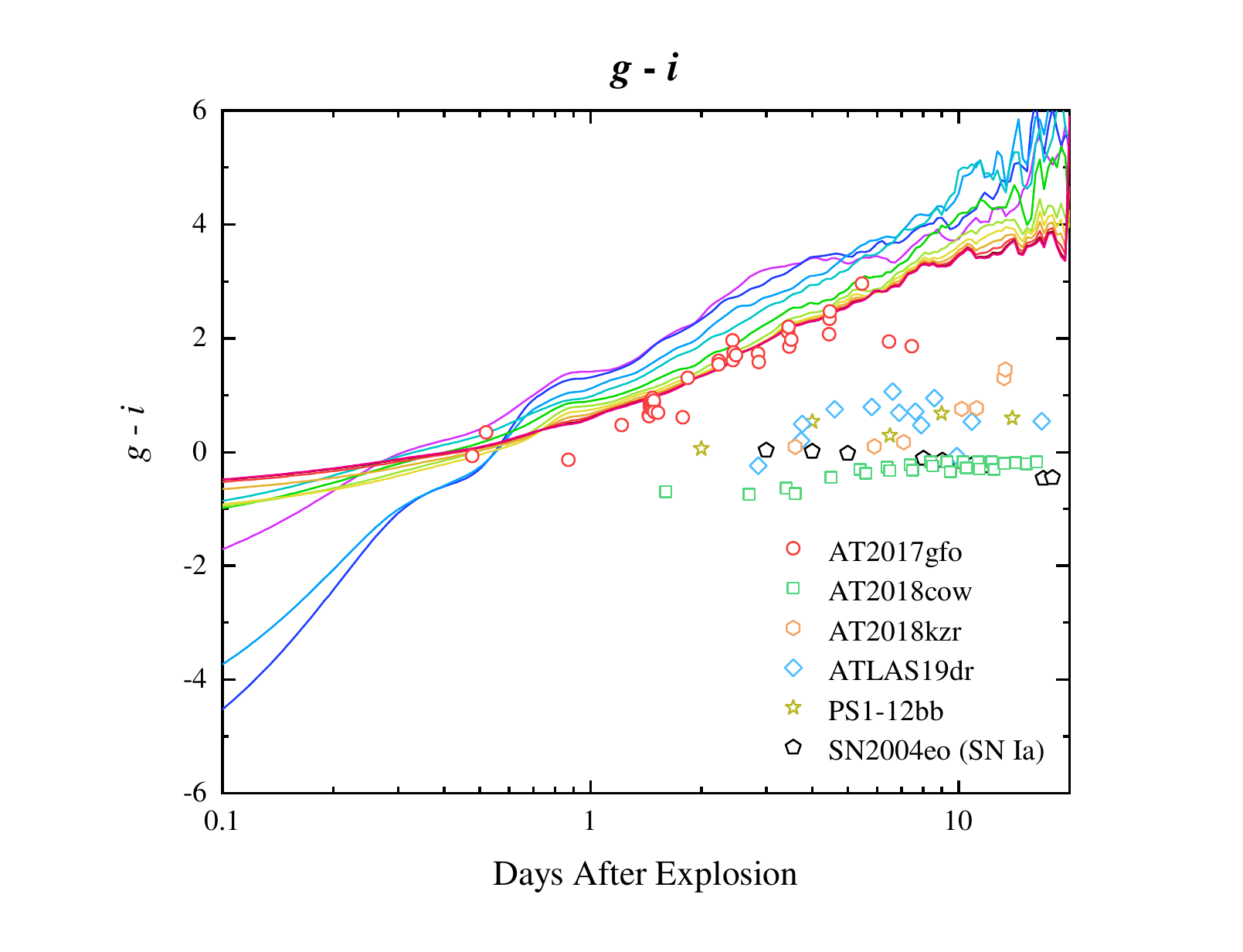}
	\includegraphics[width = 0.344\linewidth , trim = 72 20 40 60, clip]{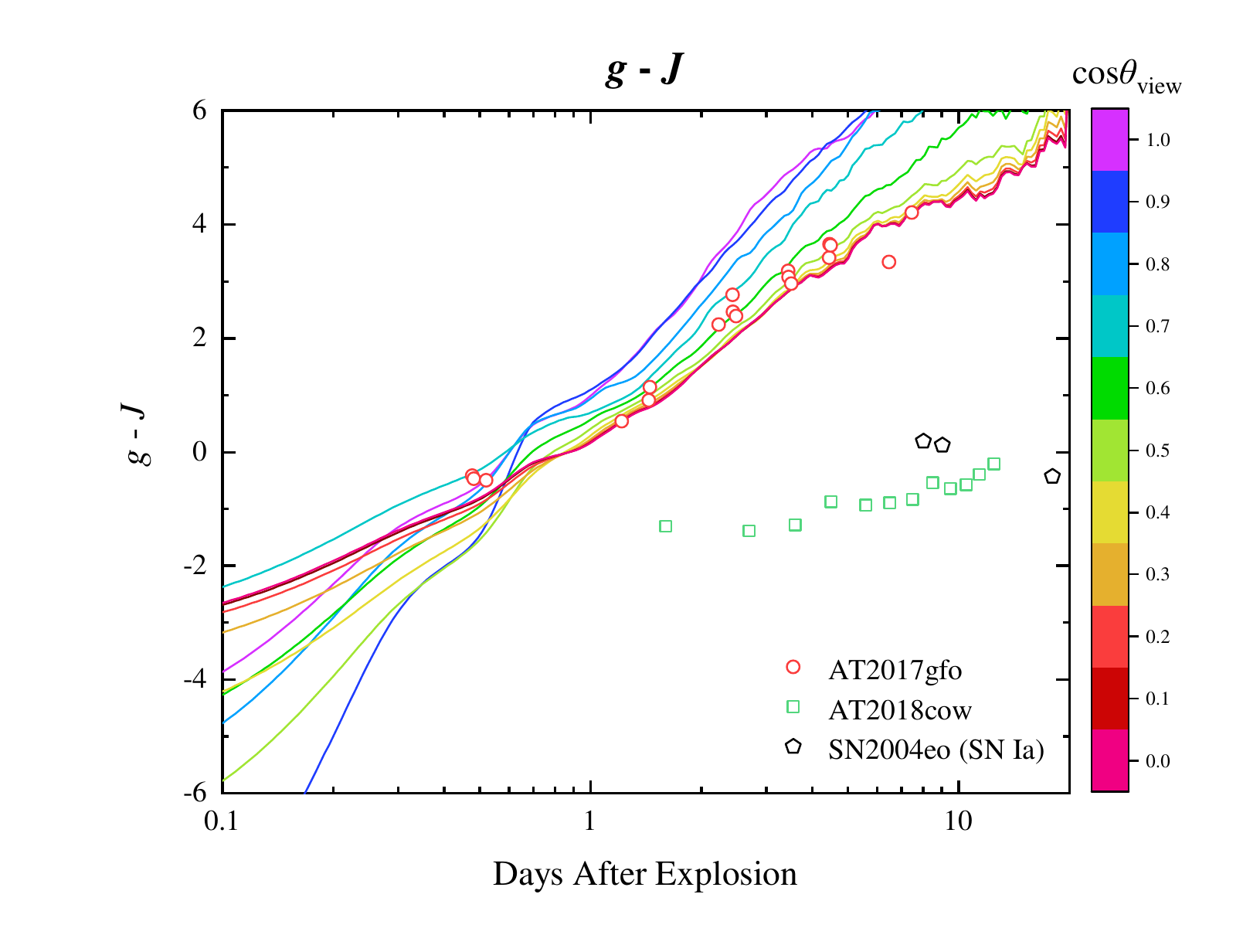}
	\includegraphics[width = 0.318\linewidth , trim = 72 30 90 60, clip]{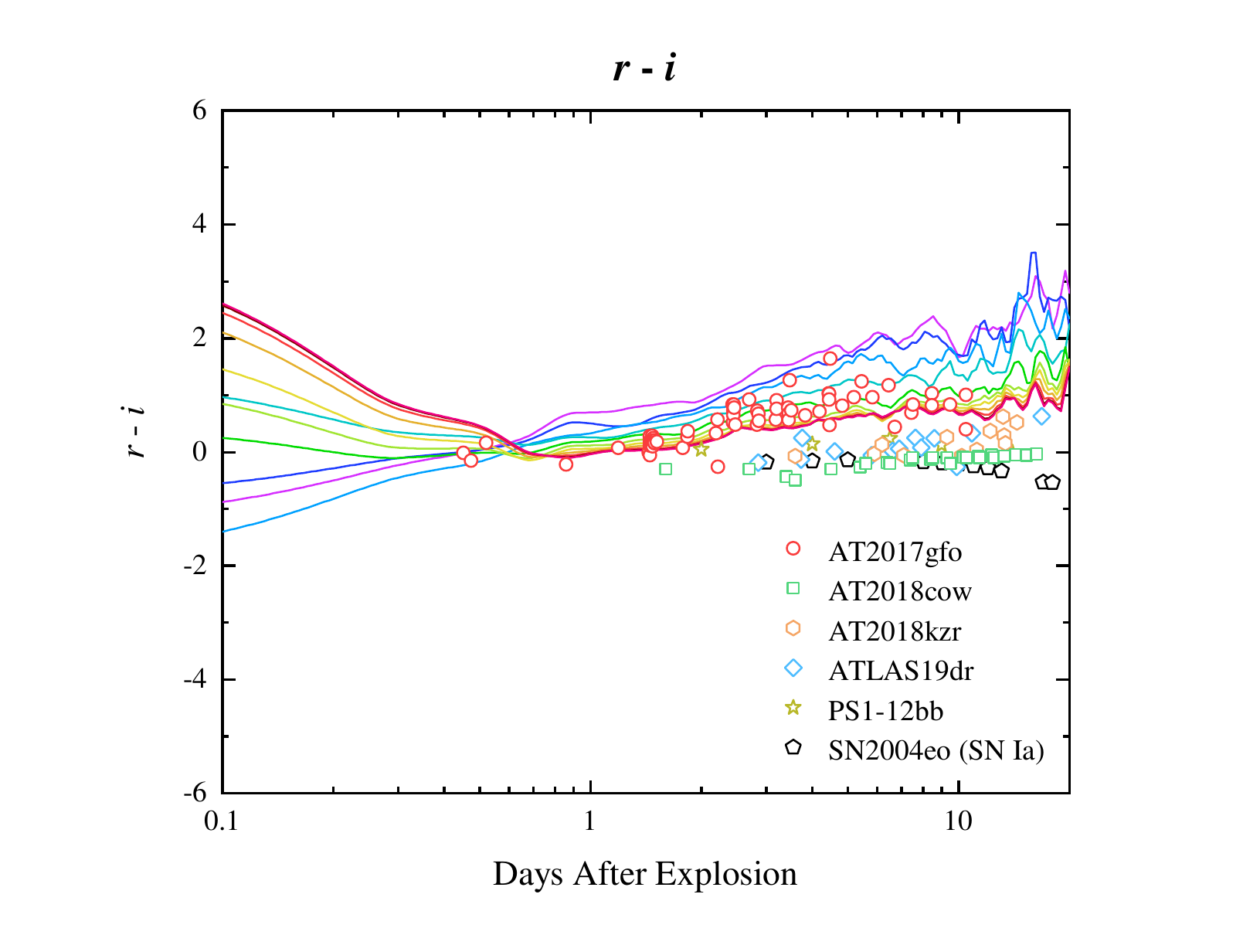}
	\includegraphics[width = 0.318\linewidth , trim = 72 30 90 60, clip]{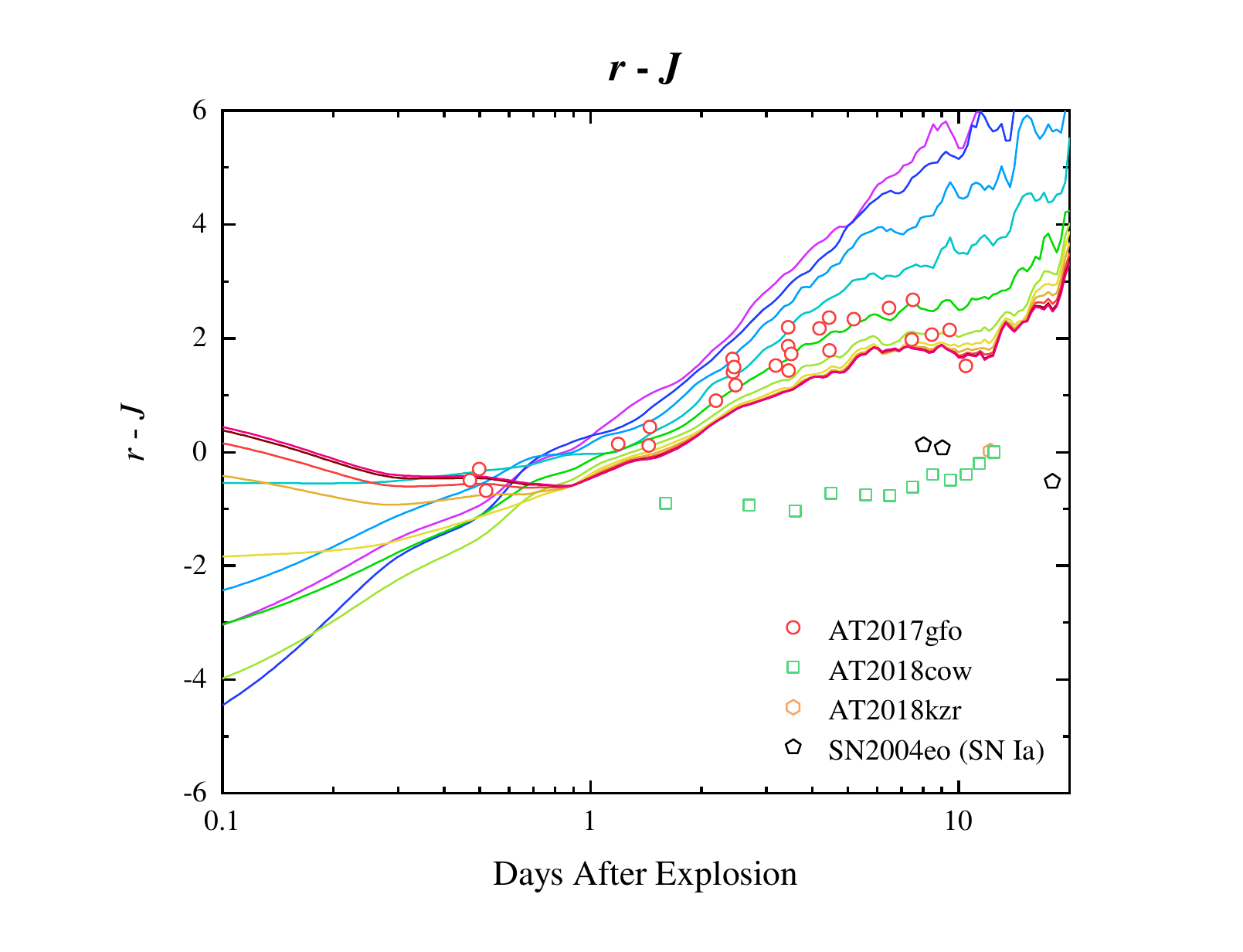}
	\includegraphics[width = 0.344\linewidth , trim = 72 30 40 60, clip]{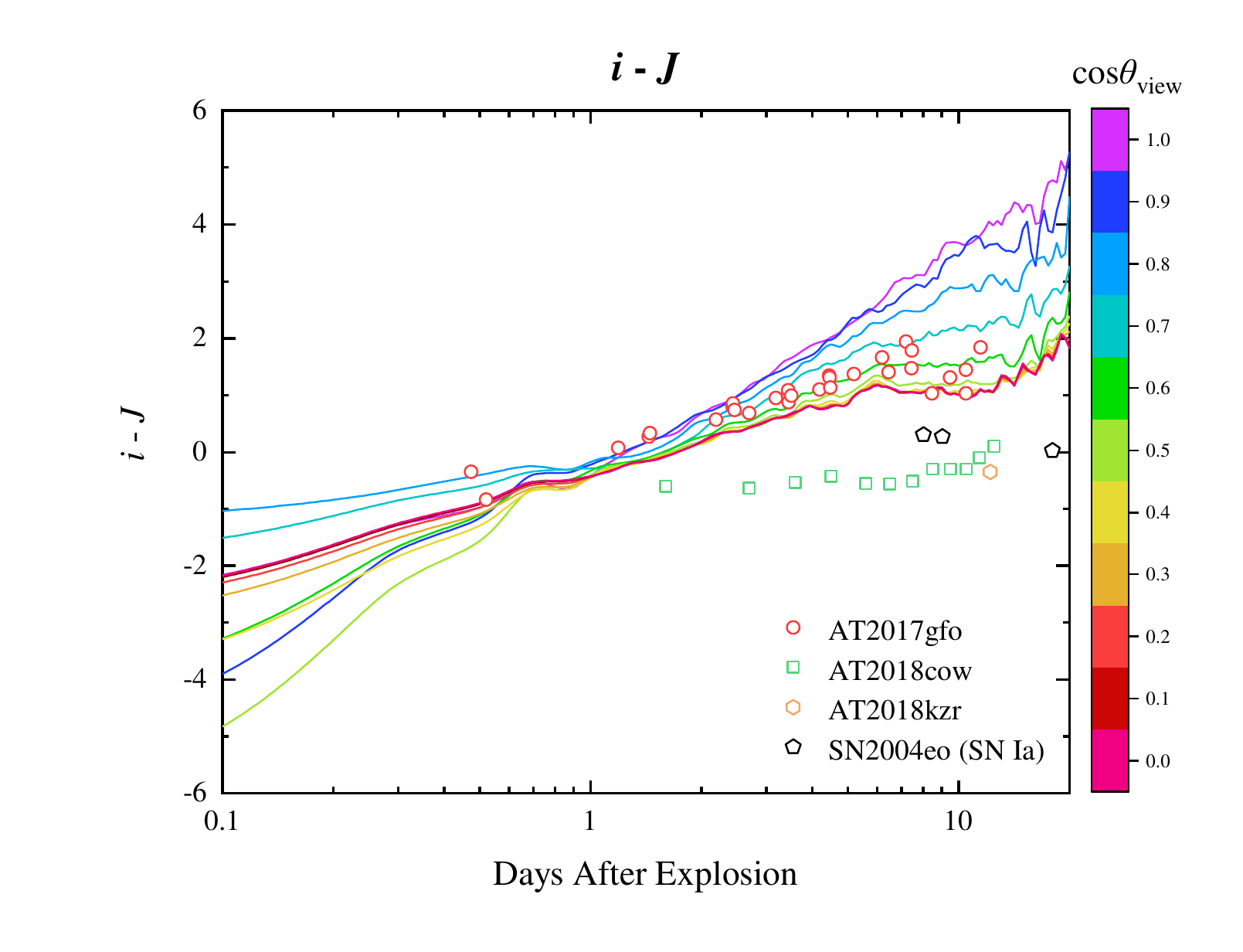}
    \caption{Color evolution of AT2017gfo (red circles), {AT2018cow} (green squares), AT2018kzr (orange hexagons), ATLAS19dqr (blue diamonds), PS1-12bb \citep[yellow stars;][]{drout2014}, and SN2004eo (SN Ia, black pentagons), in $g-r$, $g-i$, $g-J$, $r-i$, $r-J$, and $i-J$ color--{time} phase spaces. Different color lines represent the color evolution of GW170817-like kilonova modeling with different viewing angle $\sin\theta_{\rm v}$. The corresponding values of $\sin\theta_{\rm v}$ for different color lines are shown in the color bar. }\label{fig7}
\end{figure*}

Next, We discuss color evolution of AT2017gfo and other optical transients. In Figure \ref{fig7}, we plot the color evolution patterns in the $g-r$, $g-i$, $g-J$, $r-i$, $r-J$ and $i-J$ color--{time} spaces. The data points with different colors correspond to different transients, and the lines are given by the kilonova model for different viewing angles. As shown in Figure \ref{fig7}, these optical transients except AT2017gfo have a relatively small color difference between two bands (which is always smaller than $\sim 0.5\,{\rm mag}$) and a slow-evolving color evolution. The color evolution of kilonova is unique compared with the color evolution patterns of other transients. For two colors with a large difference between the high-frequency and low-frequency bands, the difference increases with time for kilonova emission. The color evolution is usually more significant between an optical band and an infrared band. {Especially for the $g-i$, $g-J$ and $r-J$ color--{time} plots, the rate of color evolution of GW170817 a few day after the BNS merger is much larger than those of other optical transients.} The unique characteristic of kilonova color evolution may be caused by the presence of multiple components with different opacities and fast-evolving changes of the opacity with temperature \citep[e.g.,][]{tanaka2020}. {By overplotting other examples of SNe and fast-evolving transients on the plot that have optical and infrared follow-up observations to resemble kilonova evolution timescales (i.e., $\sim10\,{\rm days}$),} we conclude that at least two detection epochs in two different exposure filters, {especially between an optical band and an infrared band}, could be used to directly search for and identify kilonovae in optical surveys. 

\begin{figure*}[htbp]
    \centering
	\includegraphics[width = 0.49\linewidth , trim = 10 30 20 30, clip]{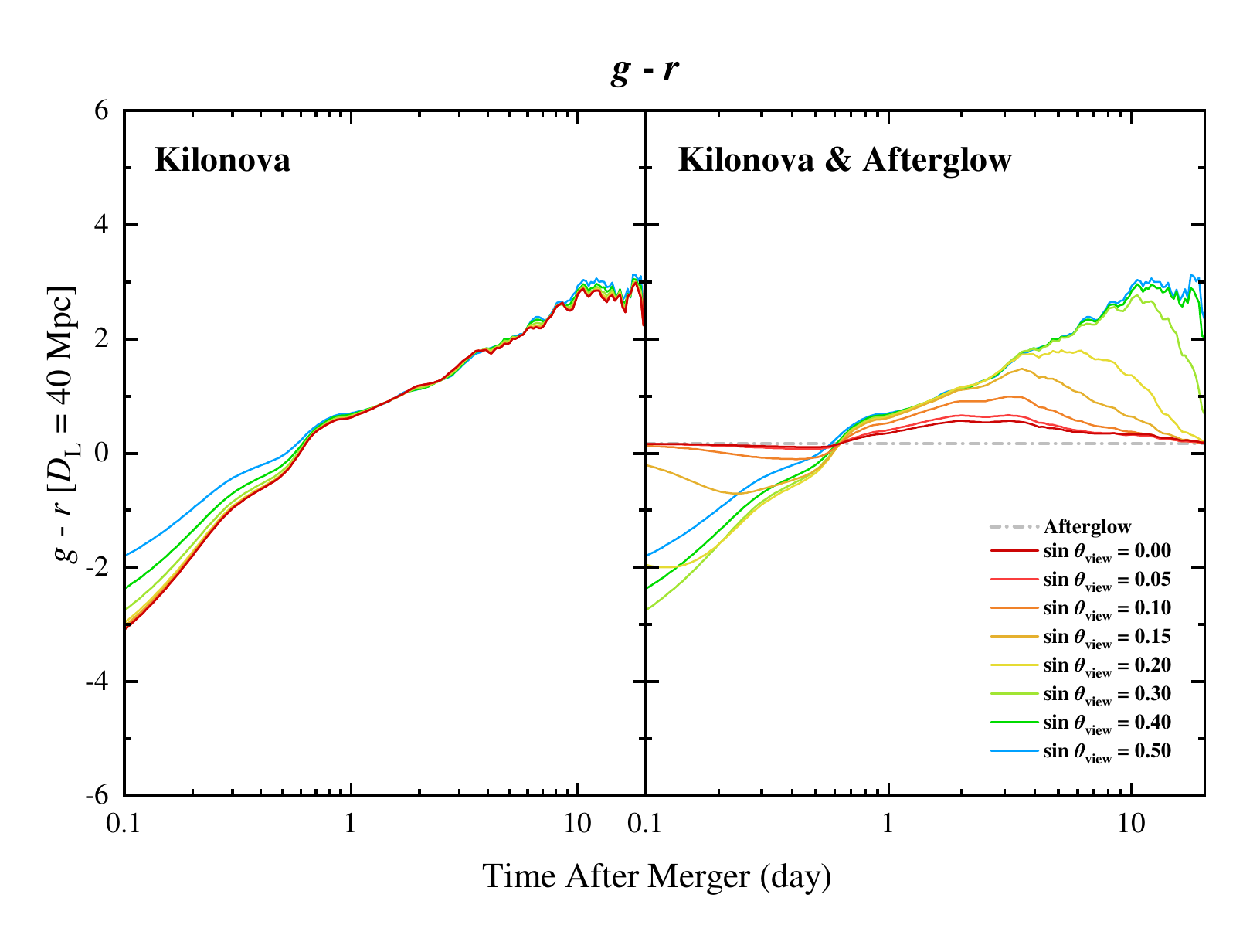}
	\includegraphics[width = 0.49\linewidth , trim = 10 30 20 30, clip]{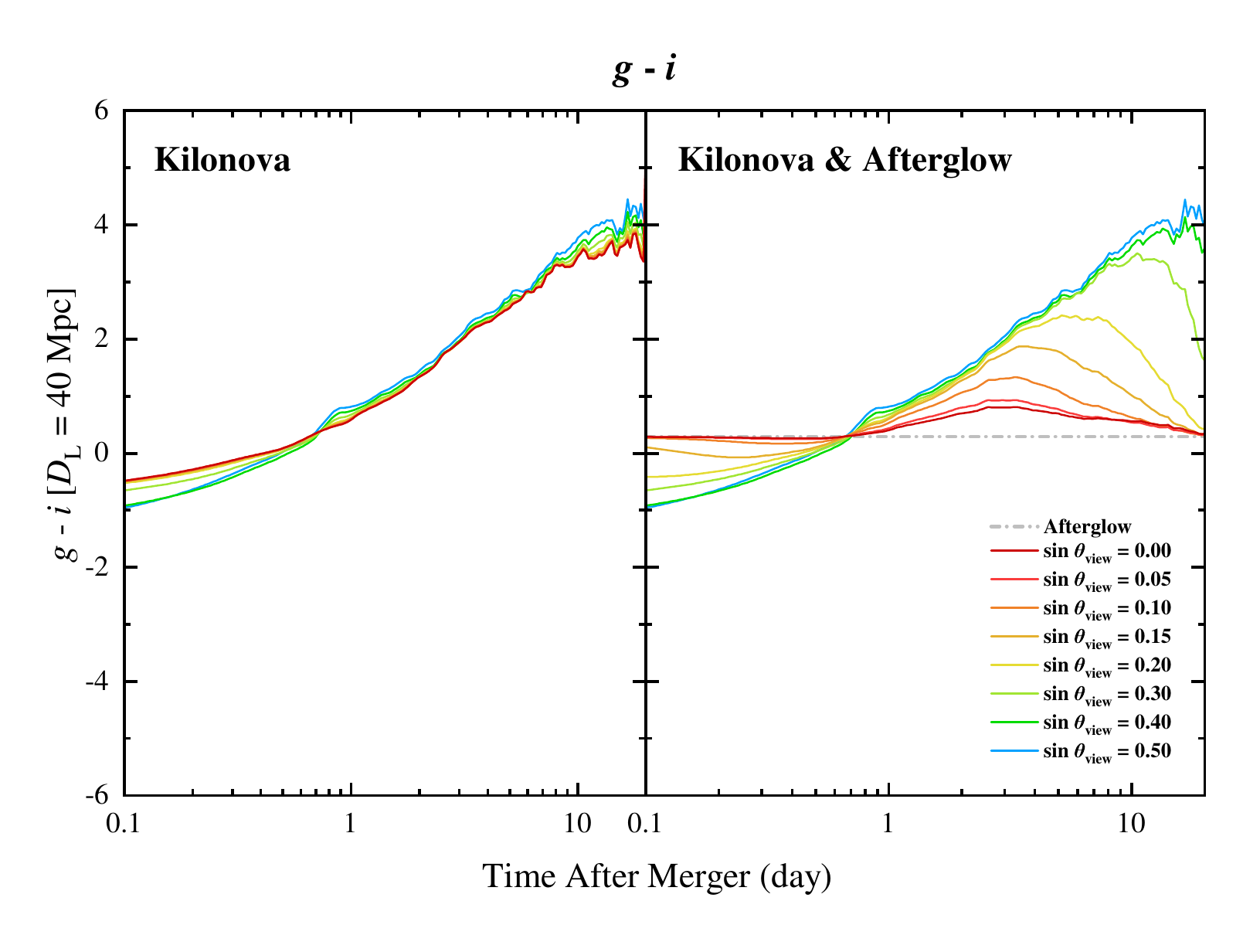}
	\includegraphics[width = 0.49\linewidth , trim = 10 30 20 30, clip]{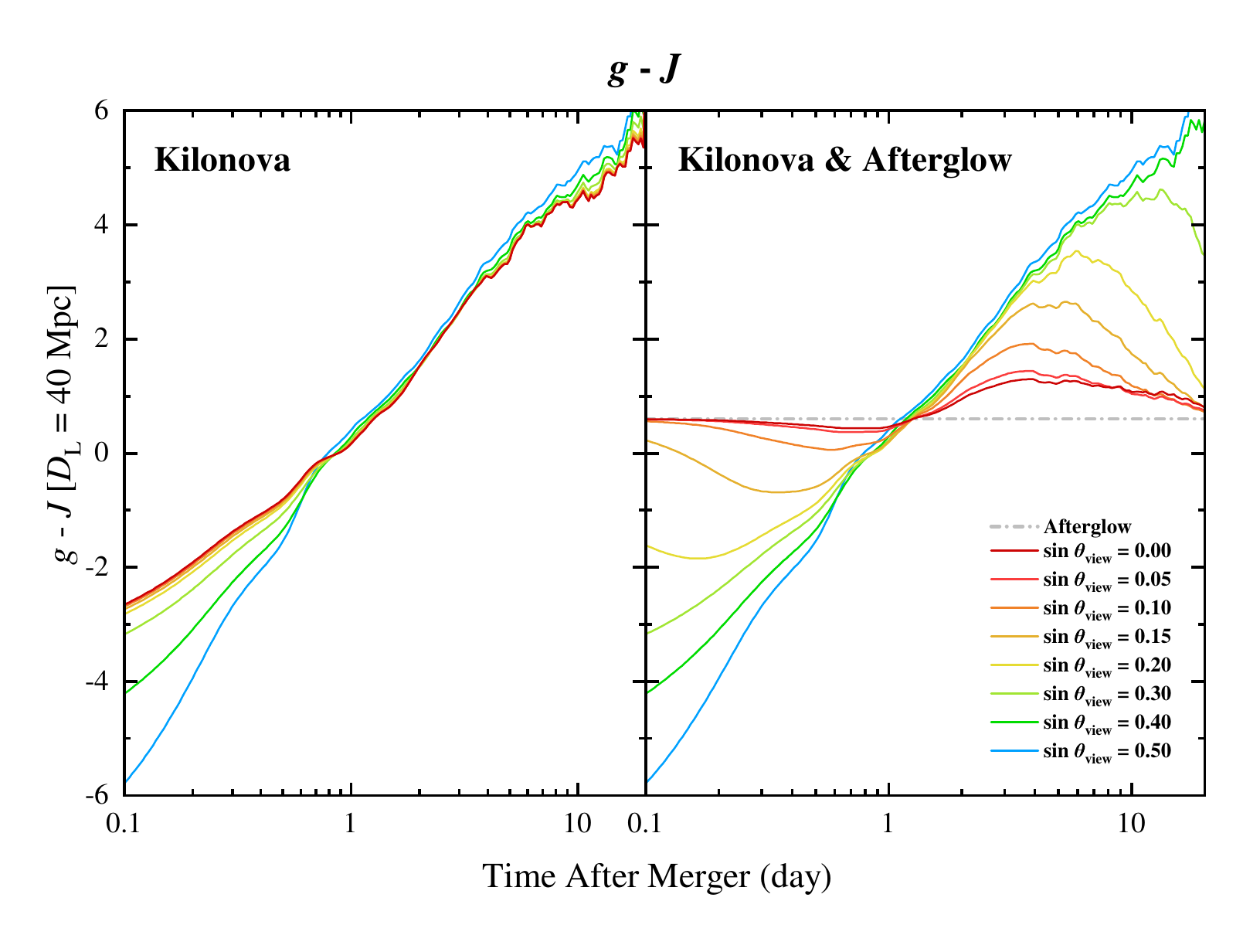}
	\includegraphics[width = 0.49\linewidth , trim = 10 30 20 30, clip]{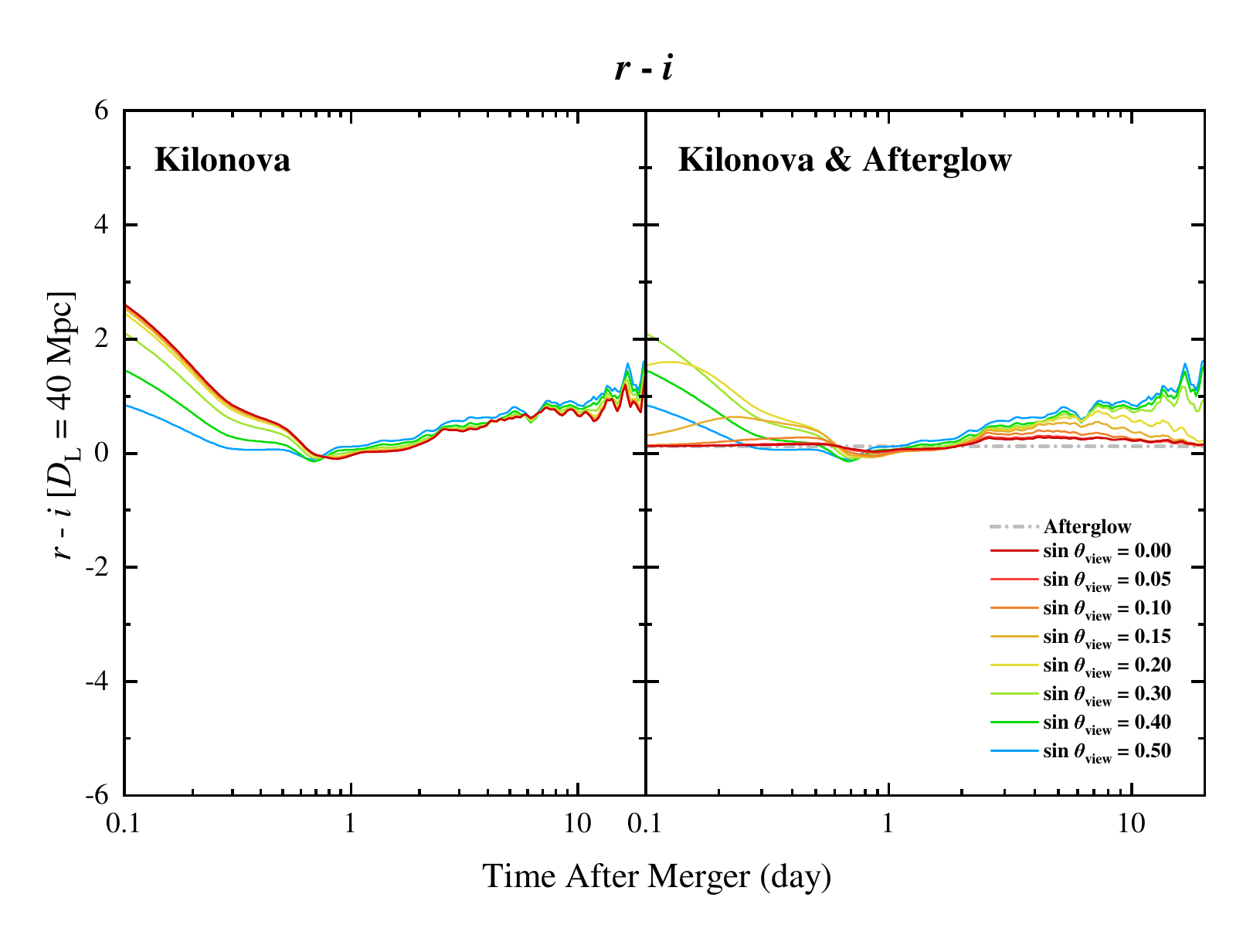}
	\includegraphics[width = 0.49\linewidth , trim = 10 30 20 30, clip]{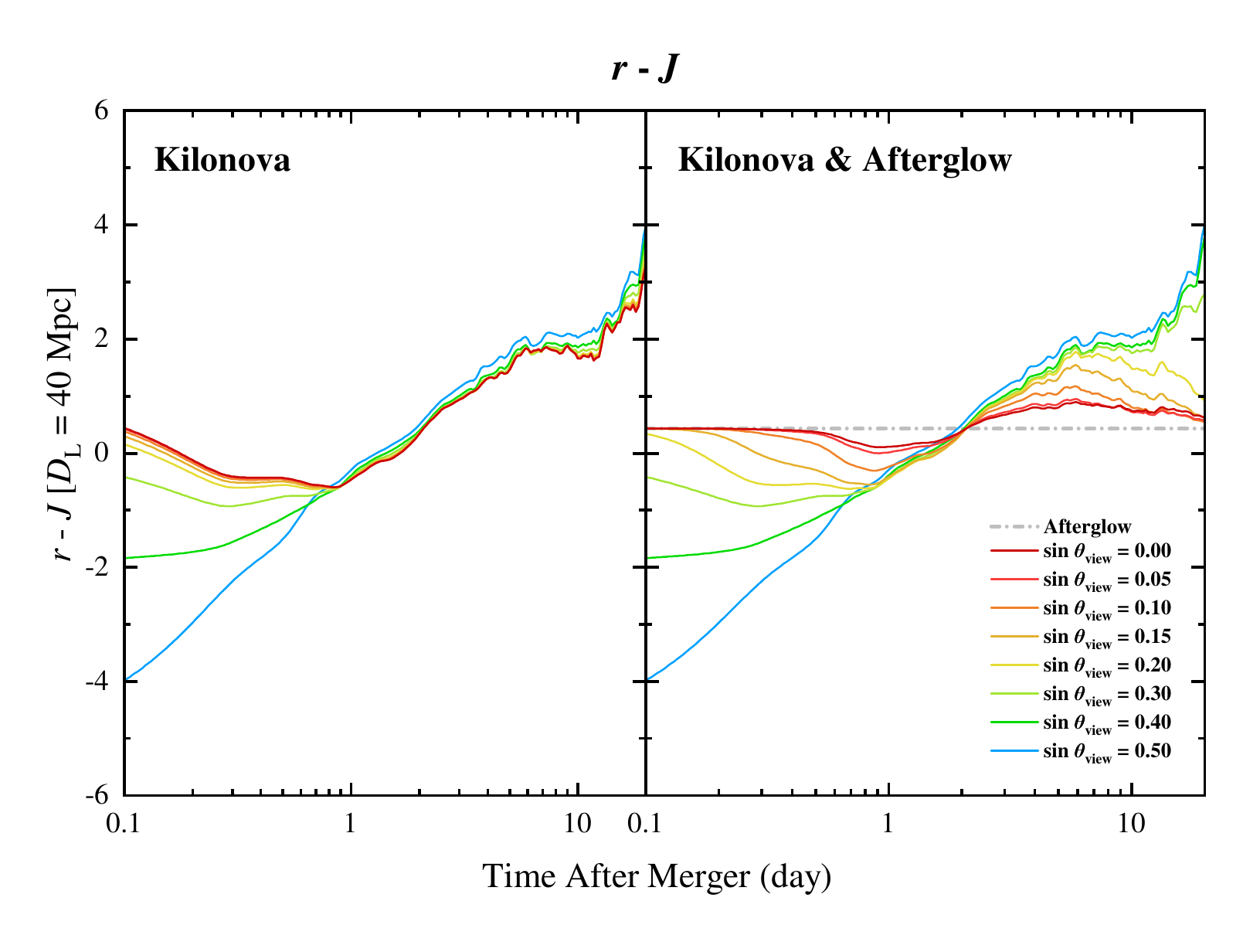}
	\includegraphics[width = 0.49\linewidth , trim = 10 30 20 30, clip]{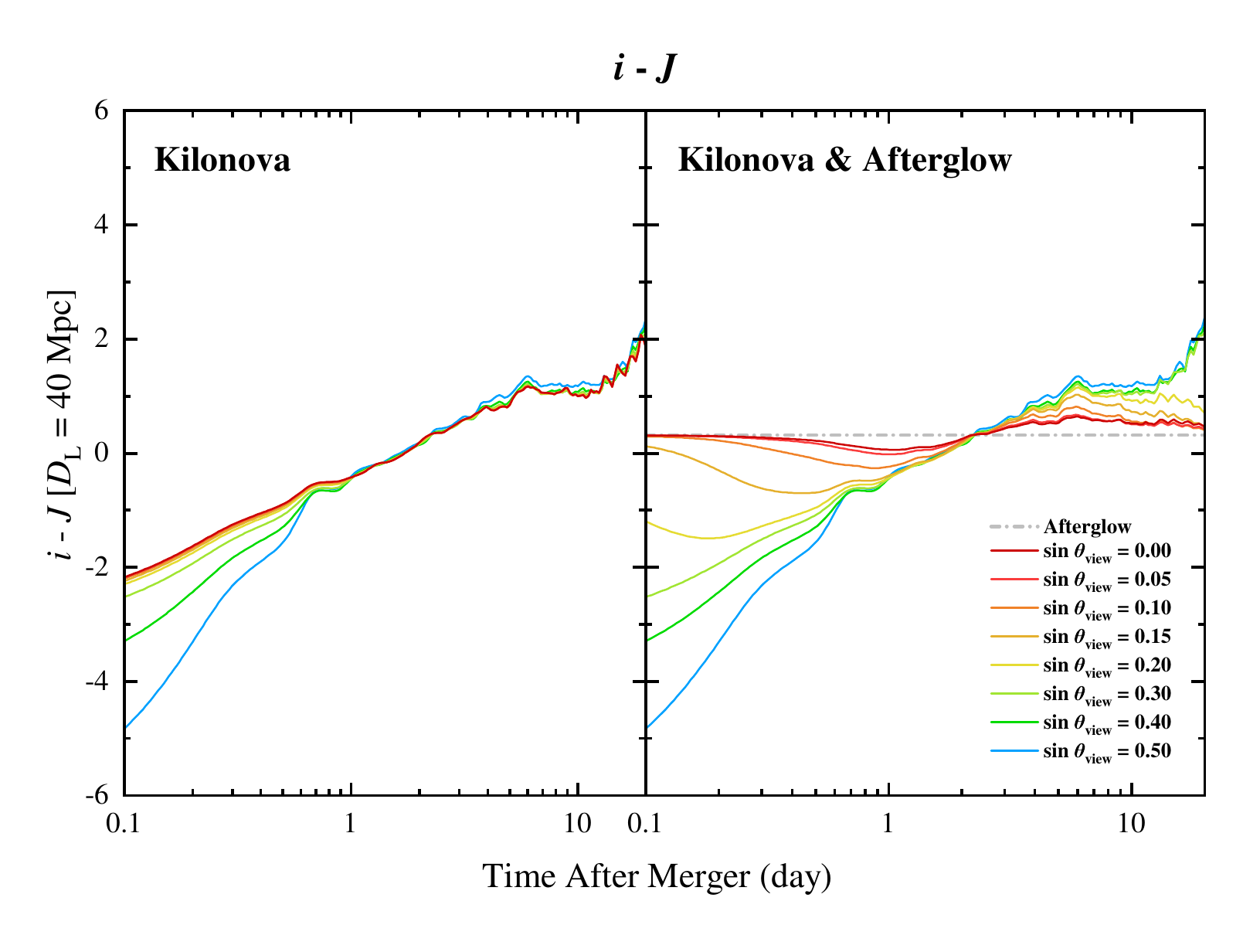}
    \caption{Color evolution patterns of our modeled GW170817-like kilonova (left sub-panels) and total emission of GW170817-like kilonova plus afterglow (right sub-panels) in $g-r$, $g-i$, $g-J$, $r-i$, $r-J$, and $i-J$ color--{time} spaces. The gray dashed-dotted lines represent  color evolution of the afterglow. The corresponding values of $\sin\theta_{\rm v}$ for different color lines are shown in the label of each panel.}\label{fig8}
\end{figure*}

We plot the color evolution patterns of our modeled kilonova and the total emission of kilonova plus afterglow for different viewing angles, as shown in Figure \ref{fig8}. In general, color evolution of afterglow is constant, because its radiation is non-thermal, but a kilonova has significant color evolution. \footnote{{We notice that long GRB (lGRB) afterglows and GRB-associated SNe from collapsars could also be directly discovered by optical serendipitous observations \citep[e.g.,][]{andreoni2021b}. The color evolution of lGRB afterglows would be constant as well, similar to that of sGRB afterglows. However, the GRB-associated SNe would emerge on the top the lGRB afterglow lightcurves much later than kilonovae emerged from sGRB afterglow lightcurves, so that GRB-associated SNe and kilonovae can be easily distinguished.}} For the on-axis case, since the optical flux is dominated by afterglow at early times, the total color evolution is approximately constant, and only appears as a small evolution a few days after the explosion when the kilonova emission becomes significant. As the viewing angle becomes larger, the total color evolution would be dominated by the kilonova, and evolve significantly. The color evolution becomes more significant when the difference between two optical bands is larger. For some detected (near) on-axis afterglow events, one can use their color evolutions to test the existence of the associated kilonovae, when the kilonova contribution is small in the observed light curve. 

\section{Summary and Conclusion}\label{sec:5}

We have carried out a detailed simulation to explore the viewing-angle-dependent properties of both kilonova and afterglow from BNS mergers. By assuming that all kilonovae are similar to AT2017gfo but viewed from different viewing angles, we explore the detectability of kilonovae in comparison with sGRB afterglows from BNSs. We found that for an on-axis view, the brightness of the afterglow at the early and late stages is always larger than that of the associated kilonova. Only about a half of on-axis afterglows are dimmer than the associated kilonovae at the kilonova peak time, in which case the kilonova could be identified as a fast-evolving bump on the afterglow lightcurve. The best detection period for the on-axis view to search for the potential kilonova signal from an afterglow is a few days after the BNS merger. Infrared bands are better to search and identify kilonovae from the afterglow lightcurves since the kilonova emission lasts longer in the infrared band. Only if $\sin\theta_{\rm v}\gtrsim0.20$, the EM signals of most BNS mergers would be kilonova-dominated. In this case, the off-axis afterglow would emerge at $\sim5-10\,{\rm day}$ after the BNS mergers. 

We simulated the kilonova and afterglow luminosity functions at the kilonova peak time. The absolute magnitude distribution of kilonova peaks at $\sim-16\,{\rm mag}$ but spans by $\sim 3\,{\rm mag}$ in optical and $\sim1-2\,{\rm mag}$ in infrared. The spans are caused by the anisotropic distribution of different ejecta components. We notice that the properties of cosmological kilonovae are assumed to be GW170817-like. However, the mass of the kilonova ejecta, especially for polar-dominated lanthanide-free neutrino-driven wind ejecta which may determine the peak luminosity of the kilonova, could be dependent on the survival time of the remnant NS \citep[e.g.,][]{kasen2015,fujibayashi201,gill2019}. Besides, the possible additional energy injection from the remnant BH via fall-back accretion \citep{rosswog2007} or the Blandford–Payne mechanism \citep{ma2018}, or from a long-lived NS through magnetic spindown \citep{yu2013,metzger2014} may enhance the brightness of the kilonova. Because the BNS kilonova emission may display significant diversity and NSBH kilonovae will also contribute to the kilonova magnitude distribution \citep{zhu2021kilonova}, it is expected that cosmological kilonova luminosity function should be more complicated than our simulation result. We also showed that the afterglows have a wider magnitude distribution compared with kilonovae. A small fraction $(\sim10\%)$ of afterglow would be brighter than kilonova at the kilonova peak time so that most BNS mergers should be kilonova-dominated. The possible presence of bright afterglow emission may significantly affect the detectability of nearby kilonovae.

After involving the cosmological redshift distribution of BNS mergers, we showed that the maximum possible apparent magnitude for cosmological kilonovae is $\sim27-28\,{\rm mag}$. The distribution range of the apparent magnitude for afterglows is wider than that of kilonovae. For the brightness that is dimmer than $\sim23-24{\rm mag}$ the afterglow luminosity function is much higher than that of kilonova. Despite the EM signatures from most of BNS mergers along the line of sight are kilonova-dominated, our result showed that the most expected EM signal from BNS mergers by serendipitous observations is afterglow rather than kilonova since the search depth for almost all the present and foreseeable future survey projects is shallower than $\sim22-23\,{\rm mag}$. It can be predicted that the number of afterglow event detected via serendipitous observations would be much higher than that of kilonova events. Thanks to the improved cadence, the ZTF has discovered a few independent optically-discovered SGRB afterglows without any detection of kilonova candidate. These observation results reported by \cite{andreoni2021} might support our calculations. {It is expected that the detection rate of kilonovae could have the same order of magnitude with that of afterglows via serendipitous observations in the era of Mephisto, WFST, LSST, etc, thanks to their unprecedented improvements of survey detectabilities. }

Since kilonovae and afterglows evolve rapidly, some candidates may be recorded in the  database of survey projects without prompt detection. We found that it may be difficult to use the fading rate in a single band to directly identify kilonovae and afterglows among various fast-evolving transients by serendipitous observations, especially if the observations have missed the peak. However, {by overplotting other examples of SNe and fast-evolving transients on the plot that have optical and infrared follow-up observations to resemble kilonova evolution timescales, we find that} one can use color evolution to identify kilonovae and afterglows thanks to their unique color evolution patterns compared with those of other fast-evolving transients. Color evolution {for kilonova emissions} is usually more significant between an optical and an infrared band, {e.g., $g-i$, $g-J$ and $r-J$}. Thus, we show that at least two detection epochs in two different exposure filters {(especially between an optical and an infrared band)} may be used to directly search for and identify kilonova and afterglows during surveys.

{\cite{lamb2017,lamb2018transient} explored the detectablility of sGRBs afterglows by considering different jet structures. Their results for the detection rate of sGRBs afterglows were obtained by comparing the distribution of peak afterglows and the threshold of survey projects. \cite{cowperthwaite2019,andreoni2021b,frostig2021,andreoni2022} focused on the studies of the search strategy and the detection rates for BNS or/and NSBH kilonovae by GW-triggered target-of-opportunity observations. \cite{scolnic2018} considered a few current and upcoming survey projects to estimate the detection rate of kilonovae via serendipitous observations by adopting specific cadence library for each survey project. Their trigger requirement was set as ``at least two filter bands have at least one observation with S/N$>5$''. However, since most kilonovae may be possible recorded with only one observation \citep[e.g.,][]{zhu2021kilonova,almualla2021} and hence, would lack the color evolution information, this requirement may be difficult to reject other fast-evolving transients. \cite{saguescarracedo2021,andreoni2021,almualla2021} adopted the pipeline of  \texttt{ZTFReST} to search for kilonovae and other fast-evolving transients through serendipitous observations. However, as shown in Figure \ref{fig6}, the selection requirement of the pipeline (fading rates faster than $0.3\,{\rm mag}\,{\rm day^{-1}}$) may hardly directly identify kilonovae from other fast-evolving transients in the survey dataset. \cite{zhu2021kilonova} explored the epochs of re-visits and the detection rate of NSBH kilonovae and afterglows by both serendipitous observations and GW-triggered target-of-opportunity observations for various survey projects. } Many works in the literature have detailedly discussed the search strategy and detection rate of {BNS} kilonova, but without the consideration of the effect of the associated afterglow emission. In \citetalias{zhu2021kilonovaafterglow}, we will investigate in detail the optical search strategy and detection rate of both kilonova and optical afterglow events from BNS mergers.

\software{\texttt{POSSIS} \citep{bulla2019,coughlin2020}; \texttt{Matlab}, \url{https://www.mathworks.com}; \texttt{Python}, \url{https://www.python.org}}

\acknowledgments

{We thank an anonymous referee for valuable suggestions and Xiao-Wei Liu, Ya-Cheng Kang for helpful comments.} J.P.Z is partially supported by the National Science Foundation of China under Grant No. 11721303 and the National Basic Research Program of China under grant No. 2014CB845800. Y.P.Y is supported by National Natural Science Foundation of China grant No. 12003028, and China Manned Spaced Project (CMS-CSST-2021-B11). H.G. is supported by the National Natural Science Foundation of China under Grant No. 11690024, 12021003, 11633001. Y.W.Y is supported by the National Natural Science Foundation of China under Grant No. 11822302, 11833003.

\bibliography{ms}{}
\bibliographystyle{aasjournal}

\end{document}